\RequirePackage{fix-cm}
\documentclass[smallextended]{svjour3}       
\smartqed  
\usepackage{graphicx}
\usepackage{amsmath}
\usepackage{amsfonts}
\usepackage{bm}
\usepackage{bbm}
\usepackage{comment}
\usepackage{hyperref}
\usepackage{soul,color}
\usepackage[titletoc]{appendix}
\usepackage[square,sort&compress,comma,numbers]{natbib} 
\usepackage{todonotes}
\usepackage{xcolor}

\newcommand{\mydrafttext}{}
\newcommand{\drafttext}[1]{\renewcommand{\mydrafttext}{#1}}
\newboolean{draft}  
\setboolean{draft}{true}
\ifthenelse{\boolean{draft}}
{
    \newcounter{comments}
    \drafttext{{\color{red}This document is a draft.}}
    \newcommand{\shane}[1]{\addtocounter{comments}{1}{\color{red}[Shane comment \thecomments: #1]}}
    \newcommand{\mattia}[1]{\addtocounter{comments}{1}{\color{blue}[Mattia comment \thecomments: #1]}}
    \newcommand{\peter}[1]{\addtocounter{comments}{1}{\color{orange}[Peter comment \thecomments: #1]}}
}
{
\newcommand{\shane}[1]{}
\newcommand{\mattia}[1]{}
\newcommand{\peter}[1]{}
}

\journalname{Revision 1}
\begin{document}

\title{Finite-time Lyapunov exponents in the instantaneous limit and material transport}

\titlerunning{FTLE in the instantaneous limit and material transport}
\authorrunning{Nolan, Serra, and Ross} 

\author{Peter J. Nolan \and Mattia Serra \and \\
        Shane D. Ross
}

\institute{Peter J. Nolan (Corresponding author) \at
              Engineering Mechanics Program,
              Virginia Tech, Blacksburg, VA 24061, USA\\
              \email{pnolan86@vt.edu}    \\      
              \\
              Mattia Serra \at 
              School of Engineering and Applied Sciences, 
              Harvard University, Cambridge, MA 02138, USA\\
              \email{serram@seas.harvard.edu} \\
              \\
              Shane D. Ross \at
              Aerospace and Ocean Engineering, Virginia Tech,
              Blacksburg, VA 24061, USA \\
              \email{sdross@vt.edu}
}

\date{This version: \today}


\maketitle

\begin{abstract}
Lagrangian techniques, such as the finite-time Lyapunov exponent (FTLE) and hyperbolic Lagrangian coherent structures (LCS), have become popular tools for analyzing unsteady fluid flows. These techniques identify regions where particles transported by a flow will converge to and diverge from over a finite-time interval, even in a divergence-free flow.  
Lagrangian analyses, however, are time consuming and computationally expensive, hence unsuitable for quickly assessing short-term material transport. A recently developed method called OECSs [Serra, M. and  Haller, G., ``Objective Eulerian Coherent Structures'', {\it Chaos} 26(5), 2016] rigorously connected Eulerian quantities to short-term Lagrangian transport.
This Eulerian method is faster and less expensive to compute than its Lagrangian counterparts, and needs only a single snapshot of a velocity field. Along the same line, here we define the instantaneous Lyapunov Exponent (iLE), the instantaneous counterpart of the FTLE, and connect the Taylor series expansion of the right Cauchy-Green deformation tensor to the infinitesimal integration time limit of the FTLE. 
We illustrate our results on geophysical fluid flows from numerical models as well as analytical flows, and demonstrate the efficacy of attracting and repelling instantaneous Lyapunov exponent structures in predicting short-term material transport.

\end{abstract}

\section{Introduction}

Lagrangian methods, such as the finite-time Lyapunov exponent (FTLE) and Lagrangian coherent structures (LCSs), have become a popular means of analyzing the Lagrangian transport structure of unsteady fluid flows and other dynamical systems \cite{shadden2005definition,lekien2007lagrangian,lekien2010computation,brunton2010fast,lipinski2010ridge,ross2010detecting,senatore2011detection,leung2011eulerian,haller2011variational,tallapagada2011lagrangian,schindler2012ridge,schindler2012lagrangian,peng2012attracting,tallapragada2013set,bozorgmagham2013real,michini2014robotic,garaboa2015lagrangian,bozorgmagham2015atmospheric,schmale2015highways,bozorgmagham2015local,haller2015review,mease2016characterizing,hadjighasem2017critical,onozaki2017tube,garaboa2017influence,schmale2017high,you2017eulerian,serra2017uncovering,chang2018transport,liu2018gulf,hsieh2018small,balasuriya2018generalized}. 
These predict the dominant particle deformation patterns in a fluid flow over a time interval of interest, as well as which regions of the flow will undergo the greatest and least amounts of stretching. However, Lagrangian methods rely on the numerical integration of particle trajectories, making such methods computationally expensive and time consuming. Furthermore, the integration of particle trajectories requires a velocity field which is sufficiently resolved in both time and space in order to accurately calculate the particle's motion. 
Beyond the issues of spatiotemporal resolution and computational time, there are applications in which one is interested in understanding short-time material transport because past or future velocities are not available or there is
no  obvious choice of the
time horizon for considering particle motion, i.e., the 
integration time $T$.
This limits the ability of researchers to compute Lagrangian diagnostics from experimental or observational data, such as from particle image velocimetry (PIV) in laboratory-scale experimental fluid mechanics \cite{shadden2006lagrangian,olcay2010sensitivity,raben2014computation,raben2014experimental}, biological applications \cite{tanaka2009separatrices,tanaka2009evaluation,tanaka2010mathematical,ross2010detecting}, such as cardiovascular flows \cite{shadden2008characterization,shadden2010computational,toger2012vortex,shadden2015lagrangian,arzani2017wall}, or from geophysical data, such as ocean currents \cite{shadden2009correlation,rypina2011near,kirincich2016remote,Dasaro2018ocean} or wind measurements \cite{nolan2018coordinated,rocha2019sensing}, necessitating the use of simulation-based flow models instead. Additionally, model data takes time to generate, limiting its usefulness for real-time time-critical applications requiring an emergency response \cite{vandop1998etex,de2014relocatable,xie2019lagrangian}, such as a hazardous incident, e.g., a radioactive material leak \cite{buesseler2012fukushima}, an oil spill \cite{mezic2010new,olascoaga2012forecasting,garcia2016dynamical,allshouse2017impact}, or ocean search-and-rescue \cite{breivik2011wind,peacock2013lagrangian,serra2019search}. Furthermore, even when model data is readily available it may not be  reliable \cite{draxler1998overview,vandop1998etex,rypina2014eulerian,ameli2019transport}. 
Thus new methods of analyzing unsteady fluid flows are required, which do not depend on particle advection schemes, and could be implemented experimentally in a local spatial neighborhood
using only Eulerian information.
 
This gap has been filled by the recent variational theory of Objective Eulerian Coherent Structures (OECSs) \cite{serra2016objective}, which assess short-term material transport in two-dimensional unsteady flows using only Eulerian quantities by exploiting the connection between the Cauchy-Green and the Eulerian rate-of-strain tensor, which is the symmetric part of the velocity field gradient. 
OECSs have successfully predicted short-term transport in several geophysical flows \cite{Serra2016,serra2017efficient}, including search-and-rescue simulations in ocean field experiments \cite{serra2019search}. 
Other methods such as the trajectory divergence rate \cite{nave2018trajectory}, the attraction and repulsion rates \cite{nolan2018coordinated,nolan2019method}, and 
Eulerian material spike formation \cite{Serra2017Separation,serra2019material,Klose2019} have been subsequently developed to analyze unsteady fluid flows. 
Motivated by \cite{serra2016objective}, most of these methods are derived from the Eulerian rate-of-strain tensor.
This allows for dynamical systems to be analyzed without the need for particle trajectory integration, which reduces the amount of time and computational power necessary for analysis. 
Furthermore, being based on gradients, these methods can be calculated from measurements using as few as $n+1$ points in the neighborhood of a point in $n$ dimensional space, and at one instant in time. For example, \cite{nolan2018coordinated} calculated the attraction rate field from experimental two-dimensional environmental fluid measurements using only 3 sampling locations to estimate the velocity gradient. 

This study builds upon the work mentioned above and further explores the connection between Eulerian quantities and short-term material transport. 
In particular, as OECSs \cite{serra2016objective} are the instantaneous limit of variational LCSs \cite{haller2015review}, here we define the instantaneous Lyapunov Exponents (iLEs) as the Eulerian limits of the backward and forward-time FTLE as  integration time goes to zero.
In addition to this, higher-order Eulerian approximations to the right Cauchy-Green deformation tensor than those currently used are derived---modified Rivlin-Ericksen tensors---expanding in the integration time $T$. Using this expansion, high-order approximations are derived for both the backward-time and forward-time FTLE fields, expanding in the integration time $T$, and using techniques from matrix perturbation theory. Analytical approximations are derived for the FTLE field for well-known examples, such as the time-varying two-dimensional double-gyre and the three-dimensional ABC flow, which have previously only had their FTLE calculated using numerical particle advection schemes. Examples based on geophysical fluid simulation data are also explored; an atmospheric data set and an oceanic data set. 
We note that an experimental example has also been considered \cite{nolan2018coordinated}.

Furthermore, a new Eulerian diagnostic tool is introduced---instantaneous Lyapunov exponent structures, or iLES, which are instantaneous iLE ridges. The iLES are shown to be the limit of FTLE ridges (i.e., FTLE-LCS) as the integration time $T$ goes to $0$, and are for general $n$-dimensional dynamical systems.
Thus, iLES provide a straightforward approach to identifying the major codimension-1 hyperbolic features dominating particle (or general phase space) deformation patterns, as the same ridge detection methods used for the FTLE field can be applied to the attraction and repulsion rate fields. It is demonstrated using analytic and realistic flows that the iLES do indeed identify the important cores of particle deformation patterns over short times. 
Moreover, as in the case of OECSs, both attracting and repelling features can be determined simultaneously as they are both based on the instantaneous velocity field gradient--one need not perform two separate particle trajectory integrations, one in forward-time, the other in backward-time. The  computational savings in using only the instantaneous velocity field, and not particle trajectory integration, is  a highlight of the method, making it a candidate for use in real-time applications.   

The paper is organized as follows. Section \ref{sec:setup} sets up the notation and makes connection with previous literature. 
Section \ref{eigS_T} 
considers instantaneous approximations of the right Cauchy-Green tensor and FTLE field. 
Section \ref{sec:ilcs} derives a new Eulerian diagnostic, iLES, as the FTLE ridge in the instantaneous limit. 
In section \ref{sec:examples}, numerous examples are provided, comparing the error of the Eulerian approximation with the benchmark FTLE field (using particle advection algorithms from \cite{DuMa2010,ameli2014development}), demonstrating the effectiveness of iLES, and comparing the attraction rate field to the backward-time FTLE field. The attraction rate and backward-time FTLE were focused on due to their usefulness in predicting where particles advected by a flow will converge, making them more relevant to real-world scenarios. 
Finally, section \ref{sec:conclusion} provides conclusions and future directions.

\section{Setup and Notation}\label{sec:setup}

Consider the dynamical system,
\begin{equation}
\begin{split}
\frac{d}{dt}\mathbf{x}(t) &= \mathbf{v}(\mathbf{x}(t),t),\label{eq:theo:1}\\
\mathbf{x}_{0} &= \mathbf{x}(t_{0}),\\
\mathbf{x}\in U \subset \mathbb{R}&^{n}, \quad t \in I \subset \mathbb{R}.
\end{split}
\end{equation}
This system can be analyzed using Lagrangian (particle trajectory) methods, by first calculating the flow map, $\mathbf{x}_{0} \mapsto \mathbf{x}_{t} =\mathbf{F}_{t_{0}}^{t}(\mathbf{x}_{0})$, for some time interval of interest, $[t_{0},t] \subset I$, where $t$ could be greater than or less than the initial time, $t_0$. The flow map, $\mathbf{F}_{t_{0}}^{t}:U \rightarrow U$, is given by,
\begin{equation}
\mathbf{F}_{t_{0}}^{t}(\mathbf{x}_{0})= \mathbf{x}_{0}+\int_{t_{0}}^{t} \mathbf{v}(\mathbf{F}_{t_{0}}^{\tau}(\mathbf{x}_{0}),\tau)\,d\tau,\label{eq:theo:3}
\end{equation}
and is typically given numerically \cite{shadden2005definition,brunton2010fast,rypina2011investigating,pratt2014chaotic} over the integration time $t-t_0$.
Taking the gradient of the flow map, $\nabla\mathbf{F}_{t_{0}}^{t}(\mathbf{x}_{0})$, the right Cauchy-Green strain tensor for the time interval of interest is, 
\begin{equation}
\mathbf{C}_{t_{0}}^{t}(\mathbf{x}_{0})= \nabla\mathbf{F}_{t_{0}}^{t}(\mathbf{x}_{0})^{\top}
\nabla\mathbf{F}_{t_{0}}^{t}(\mathbf{x}_{0}),\label{eq:theo:4}
\end{equation}
where $(^{\top})$ denotes the matrix transpose.
The right Cauchy-Green tensor $\mathbf{C}_{t_{0}}^{t}(\mathbf{x}_{0})$
physically represents the material deformation of infinitesimal volume elements, and as a matrix,
is symmetric and positive-definite, giving positive eigenvalues  which can be ordered as,
\begin{equation}
    \lambda_{1}\le\lambda_{2}\le \cdots \le \lambda_{n},
    \label{C_eigvals}
\end{equation}
with associated normalized eigenvectors,
\begin{equation}
    \bm{\xi}_{\lambda_{i}}, \quad  i\in \{1,\ldots,n\}.
    \label{C_eigvecs}
\end{equation}
From the maximum eigenvalue of the right Cauchy-Green tensor, the finite-time Lyapunov exponent (FTLE) \cite{shadden2005definition,lekien2007lagrangian} can be defined as,
\begin{equation}
    \sigma_{t_{0}}^{t}(\mathbf{x}_{0}) = \frac{1}{2|T|}\log(\lambda_{n}),
    \label{FTLE}
\end{equation}
where $T= t-t_0$ is the (signed) elapsed time, also the integration time, as mentioned above. Taking the instantaneous spatial gradient of  the velocity field $\mathbf{v}(\mathbf{x},t)$ in  \eqref{eq:theo:1}, we consider the Eulerian rate-of-strain tensor \cite{truesdell1965handbuch,weiss1991,dresselhaus_tabor_1992,koh2002,serra2016objective},
\begin{equation}
    \mathbf{S}(\textbf{x},t) \equiv \tfrac{1}{2} \Big(\nabla\mathbf{v}(\mathbf{x},t)+\nabla\mathbf{v}(\mathbf{x},t)^{\top}\Big), \label{S_matrix}
\end{equation}
which is a symmetric matrix, yielding eigenvalues which are real and can be ordered as,
\begin{equation}
    s_{1}\le s_{2}\le \cdots \le s_{n},
\end{equation}
with associated normalized eigenvectors, 
\begin{equation}
    \bm{e}_{i}, \quad  i\in \{1,\ldots,n\}.
\end{equation}
 
Taylor-expanding $\mathbf{C}_{t_{0}}^{t}(\mathbf{x}_{0})$ in $t$, \cite{serra2016objective} found that $\mathbf{S}(\textbf{x},t)$ governs, at leading order, short-time material deformation in the phase space, and then derived precise definitions for OECSs using the same concepts already developed for variational LCSs. OECSs can be of three types:\ hyperbolic (attracting and repelling), parabolic and elliptic. Hyperbolic OECSs---the relevant one in this context---are the instantaneously most attracting and repelling structures in a dynamical system, mimicking the role of stable and unstable manifolds over short times. Parabolic OECSs are short-term jet-type structures, serving as short-term pathways for material transport. Elliptic OECSs are short-term vortical structures. We now   proceed with the development of iLES, and then detail their comparison with the hyperbolic OECSs.


 
\section{Expansion of the right Cauchy-Green tensor in the infinitesimal integration time limit}\label{eigS_T}



Considering the $n$-dimensional dynamical system \eqref{eq:theo:1}, for small $|T|$, one can perform a Taylor series expansion of $\mathbf{C}_{t_{0}}^{t}(\mathbf{x})$ in $T$ as,
\begin{equation}
\mathbf{C}_{t_{0}}^{t}(\mathbf{x}) =\mathbbm{1} + 2T\mathbf{S}(\textbf{x},t_0) 
+ T^2 \mathbf{B}(\textbf{x},t_0)
+ \tfrac{1}{2}T^3\mathbf{Q}(\textbf{x},t_0) + \mathcal{O}(T^4),
\label{Cauchy_Green_expansion}
\end{equation}
where $\mathbbm{1}$ is the $n \times n$ identity and
where $\mathbf{S}$, $\mathbf{B}$, and $\mathbf{Q}$ are 
the first three Rivlin-Ericksen tensors \cite{truesdell1965handbuch}, re-scaled to put them in a form which makes \eqref{Cauchy_Green_expansion} more amenable to matrix perturbation analysis for small $|T$.  The $\mathbf{S}$ matrix was defined in \eqref{S_matrix}, and
$\mathbf{B}$ and $\mathbf{Q}$ are given by,
\begin{equation}
\mathbf{B}(\textbf{x},t_0) \equiv 
\frac{1}{2}\Big[
\nabla \mathbf{a}(\textbf{x},t_0) +
\nabla \mathbf{a}(\textbf{x},t_0)^{\top}
\Big] +
\nabla\mathbf{v}(\mathbf{x},t_0)^{\top} 
\nabla\mathbf{v}(\mathbf{x},t_0),
\label{B_matrix}
\end{equation}
where the acceleration field, $\mathbf{a}(\textbf{x},t_0)$, is,
\begin{equation}
\mathbf{a}(\textbf{x},t_0)=\frac{d}{dt}\mathbf{v}(\textbf{x},t_0) = 
\frac{\partial}{\partial t} \mathbf{v}(\textbf{x},t_0) 
+ \mathbf{v}(\textbf{x},t_0) \cdot \nabla \mathbf{v}(\textbf{x},t_0),
\label{acceleration_field}
\end{equation}
where $(\cdot)$ represents the usual dot product on $\mathbb{R}^n$.
The acceleration $\mathbf{a}(\textbf{x},t_0)$ is
the total time derivative of $\mathbf{v}(\textbf{x},t_0)$, that is, the  acceleration measured along a trajectory (i.e., in a Lagrangian frame).  The matrix $\mathbf{Q}$ 
is,
\begin{equation}\label{Qmatrix}
   \mathbf{Q} \equiv \frac{1}{3}\left[\nabla\frac{d\mathbf{a}}{dt} + \left(\nabla\frac{d\mathbf{a}}{dt}\right)^{\top}\right]
 +  \left[ 
 \left(\nabla\mathbf{v}\right)^{\top} \nabla\mathbf{a} 
 + 
 \left(\nabla\mathbf{a}\right)^{\top} \nabla\mathbf{v} 
 \right].
\end{equation}
Details are given in Appendix \ref{appendix}.

If $s_1$ (respectively, $s_n$) is denoted as $s_{-}$ ($s_{+}$), with corresponding eigenvector $\bm{e}_{-}$ ($\bm{e}_{+}$), the main result on the short-time $T<0$ ($T>0$) approximation of the backward (forward) time FTLE field can be summarized as follows, with terms through second order in $T$  included, 
\begin{equation}
\begin{split}
\sigma_{t_{0}}^{t}(\mathbf{x}) = &\pm s_{\pm} \pm a_{\pm} T \pm b_{\pm} T^2 + \mathcal{O}(T^3), \\
&{\rm for}~ {\rm sign}(T)={\rm sign}(t-t_0)=\pm 1,
 \label{FTLE_T_correction2}
 \end{split}
\end{equation}
where,
\begin{equation}
\begin{split}
a_{\pm} &= - s_{\pm}^2 + \tfrac{1}{2} \mu_{1\pm}, \\
b_{\pm} &= \tfrac{4}{3}s_{\pm}^3 -s_{\pm} \mu_{1\pm} + \tfrac{1}{4}\mu_{2\pm}, \label{a_and_b_variables}
\end{split}
\end{equation}
with, 
\begin{equation}
\begin{split}
\mu_{1\pm} &= \bm{e}_{\pm}^{\top}\mathbf{B} \bm{e}_{\pm}, \\
\mu_{2\pm} &=
\bm{e}_{\pm}^{\top}\mathbf{Q}\bm{e}_{\pm}
+\bm{e}_{\pm}^{\top}\mathbf{B} \bm{\xi}_{1\pm} - \mu_{1\pm} \bm{e}_{\pm}^{\top}\bm{\xi}_{1\pm}. 
\label{mu_variables}
\end{split}
\end{equation}
where $\bm{\xi}_{1\pm}$ is the vector solution of,
\begin{equation}
(\mathbf{S} - s_{\pm} \mathbbm{1})\bm{\xi}_{1\pm} = - (\mathbf{B} - \mu_{1\pm} \mathbbm{1}) \bm{e}_{\pm}.
\label{xi_1_main}
\end{equation}
where the dependence on $\textbf{x}$ and $t_0$ is understood, the `$-$' terms correspond to $T<0$ and the `$+$' terms correspond to $T>0$.
We refer to the approximation of the finite-time Lyapunov exponent field, based on the instantaneous velocity field $\mathbf{v}(\textbf{x},t_0)$, as the {\it instantaneous Lyapunov exponent}, or {\it iLE}.
The details are in the Appendix, 
as well as a simplified method for obtaining $\mu_{2\pm}$ in the case of two-dimensional flows which does not require calculating $\bm{\xi}_{1\pm}$.

In the infinitesimal limit, from \eqref{FTLE_T_correction2}, the iLE field is,
\begin{equation}
    \sigma_{t_{0}}^{t}(\mathbf{x}) = 
    \pm s_{\pm}(\textbf{x},t_0) \quad \text{as} \quad t-t_0 \rightarrow 0^{\pm}
\end{equation}
Note that the connection between the proportionality of the eigenvalues of $\mathbf{S}(\textbf{x},t_0)$ and the FTLE field for small $|T|$ was suggested by \cite{perez2013path}, whereas here the equality in the limit as $|T|\rightarrow 0$ is proven. Furthermore, we have provided a framework to explicitly write the expansion of the FTLE field through any order in $T$, via a straightforward extension of \eqref{Cauchy_Green_expansion} through order $k$ in $T$. 
We also note that  the eigenvectors of $\mathbf{S}(\textbf{x},t_0)$ and $\mathbf{C}_{t_{0}}^{t}(\mathbf{x})$ are equal in the limit as $T$ goes to zero (see the Appendix).

For the $n=2$ dimensional case with $\mathbf{x}=(x,y)$, with the vector field denoted $\mathbf{v}=(u,v)$, 
the instantaneous attraction and repulsion rates 
are given analytically by,
\begin{equation}
s_{\pm}(\mathbf{x},t_0) =
\tfrac{1}{2}{\rm div}(\mathbf{v}(\mathbf{x},t_0)) \pm \tfrac{1}{2}\varepsilon_{Tot}(\mathbf{x},t_0).
\label{s_analytical_explicit_fluid}
\end{equation}
in terms of commonly used fluid quantities \cite{stewart1945,saucier1953,okubo1970,chelton2011,perez2013path,vortmeyerkley2016}, 
where,
\begin{equation}
{\rm div}(\mathbf{v}(\mathbf{x},t_0)) = \nabla \cdot \mathbf{v}(\mathbf{x},t_0)=\tfrac{\partial u}{\partial x} + \tfrac{\partial v}{\partial y},
\end{equation}
is the divergence of the flow field, 
$\varepsilon_N(\mathbf{x},t_0)= \tfrac{\partial u}{\partial x} - \tfrac{\partial v}{\partial y}$ is the
normal component of the strain rate,  
$\varepsilon_S(\mathbf{x},t_0)= \tfrac{\partial u}{\partial y} + \tfrac{\partial v}{\partial x}$ is the shear component of the strain rate, and,
\begin{equation}
\varepsilon_{Tot}(\mathbf{x},t_0)=\sqrt{\varepsilon_N^2(\mathbf{x},t_0)  
+\varepsilon_S^2(\mathbf{x},t_0)  },
\end{equation}
is the total strain rate. 

For an incompressible (i.e., divergence-free) two-dimensional flow, notice 
$s_{\pm}(\mathbf{x},t_0) =\pm \tfrac{1}{2}\varepsilon_{Tot}(\mathbf{x},t_0)$, as noted by \cite{koh2002}, and thus,
\begin{equation}
\sigma_{t_{0}}^{t}(\mathbf{x}) =\pm \tfrac{1}{2}\varepsilon_{Tot}(\mathbf{x},t_0),
\label{s_analytical_explicit_fluid_incompressible}
\end{equation}
as $t-t_0 \rightarrow 0$, that is, the forward and backward FTLE are the negative of each other, hence  have the {\it same structure}, in the infinitesimal integration time limit. Similarly, for incompressible flows, attracting and repelling OECSs are perpendicular to each other and their intersection is called \textit{objective saddle point} \cite{serra2016objective}. 
While the attraction and repulsion rate fields are the same  in the infinitesimal limit (differing only by a minus sign), the corresponding eigenvector fields, $\bm{e}_{\pm}(\mathbf{x},t_0)$, need not be the same, and in fact are perpendicular almost everywhere. This has implications, as shown in an example below, for approximating the instantaneous most attracting or repelling material surfaces.  

\section{Definition of instantaneous Lyapunov exponent structures as ridges of iLE and their significance for finite-time transport}\label{sec:ilcs}
Previous work \cite{shadden2005definition,lekien2007lagrangian,schindler2012ridge,senatore2011detection,tallapagada2011lagrangian,tallapragada2013set,bozorgmagham2013real,bozorgmagham2015local,bozorgmagham2015atmospheric} has demonstrated that 
hyperbolic 
LCSs
can be identified as ridges of the FTLE field. While there are different mathematical definitions for what constitutes a ridge, a co-dimension 1 ridge can be thought of as the generalization of the concept of local maxima. For this study, hyperbolic LCSs will be identified as C-ridges of the FTLE field. C-ridges were first described in \cite{peikert2012ridge}, as ridges of the FTLE which are orthogonal to the direction of maximal stretching. They are defined as,
\begin{align}
    \sigma > 0,\\
    \nabla \sigma \cdot \bm{\xi}_{\lambda_{n}}=0,\\
    (\textbf{H}_{\sigma} \cdot \bm{\xi}_{\lambda_{n}}) \cdot \bm{\xi}_{\lambda_{n}}<0,
\end{align}
where the dependence on $\mathbf{x}$,  $t_0$, and $t$ is understood, and $\textbf{H}_\sigma$ denotes the Hessian of the FTLE field.
C-ridges are advantageous over other definitions of ridges for the FTLE field, as they only rely on invariants of the right Cauchy-Green strain tensor. 

We propose an instantaneous approximation to the traditional finite-time-FTLE-based hyperbolic LCS:
the instantaneous Lyapunov exponent structure.
Following \cite{peikert2012ridge}, we seek co-dimension 1 manifolds in the phase space which maximize local stretching and are orthogonal to the direction of maximal stretching. 
For a  flow $\mathbf{F}_{t_0}^t$, the FTLE field provides a measure of stretching over a given time period. As $-s_{-}$ and $s_{+}$ are the limits of the backward-time and forward-time FTLE fields as integration time goes to $0$, we seek ridges of these fields which are orthogonal to the direction of maximal stretching. 
A ridge of the iLE is an {\it instantaneous Lyapunov exponent structure}, or {\it iLES}. The direction of maximal stretching in a flow over a time interval is the given by the eigenvector of the right Cauchy-Green strain tensor associated with the largest eigenvalue. As the eigenvectors of the right Cauchy-Green and Eulerian tensors are equal in the infinitesimal-time limit, we seek ridges of $-s_{-}$ and $s_{+}$ which are orthogonal to their associated eigenvector. Following \cite{peikert2012ridge}, these ridges will be referred to as S-ridges. S-ridges are thus the limit of C-ridges as integration time goes to zero. 
An attracting iLES  is a ridge of $-s_-$, thus it can be defined as a trench of $s_-$, that satisfies the following conditions.

\begin{figure}[!t]
\begin{center}    
\includegraphics[width=1\linewidth]{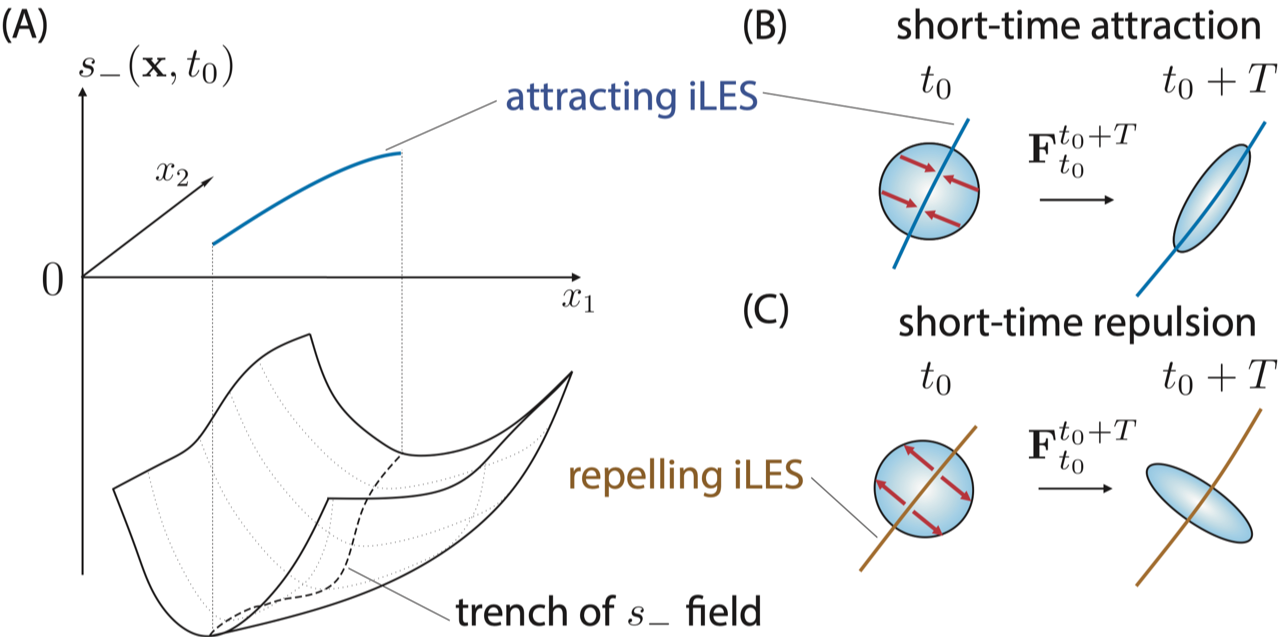}
    \caption{(A) Schematic of the attracting rate field, $s_{-}(\mathbf{x},t_0)$, at a fixed time $t_0$, showing an attracting iLES, a  trench of the $s_{-}$ field, along with (B) the short-term deformation of an infinitesimal area element (i.e., a fluid blob) centered on an attracting iLES, where $T>0$ is small. (C) A schematic of a repelling iLES and the effect on a fluid blob over small time $T>0$.}
    \label{fig:iLES_definition}
\end{center}
\end{figure}

\begin{definition} 
An {\it attracting iLES} of the system \eqref{eq:theo:1} is a co-dimension 1 manifold which satisfies,
\begin{align}
    s_{-} < 0, \label{s1_criterion} \\
    \nabla s_{-} \cdot \bm{e}_{-}=0, \label{s1_ridge_criterion} \\
    (\textbf{H}_{s_{-}} \cdot \bm{e}_{-}) \cdot \bm{e}_{-}>0. 
    \label{s1_ridge_concavity_criterion}
\end{align}
where the dependence on $\mathbf{x}$,  $t_0$, and $t$ is understood.
\end{definition}
Additionally, as a ridge of $s_+$, a repelling iLES 
is 
defined as follows.
\begin{definition} 
A {\it repelling iLES} of the system \eqref{eq:theo:1} is a co-dimension 1 manifold which satisfies,
\begin{align}
    s_{+} > 0, \label{sn_criterion}\\
    \nabla s_{+} \cdot \bm{e}_{+}=0, \label{sn_ridge_criterion}\\
    (\textbf{H}_{s_+} \cdot \bm{e}_{+}) \cdot \bm{e}_{+}<0. \label{sn_ridge_concavity_criterion}
\end{align}
where the dependence on $\mathbf{x}$,  $t_0$, and $t$ is understood.
\end{definition}
These definitions are illustrated schematically, along with the effect on a local fluid parcel, in two dimensions in Fig.\ \ref{fig:iLES_definition}.

In two-dimensional systems, OECSs and iLESs can be identified.
Attracting OECSs are one-dimensional curves ($\gamma_i,\ i\in\{1,...,N\}$) tangent to the  eigenvector $\bm{e}_{+}$ and emanating from minima of $s_{-}$, which demarcate their attracting core. The instantaneous attraction rate to $\gamma_i$ is quantified exactly by the local $s_{-}$ field \cite{serra2016objective}. 
Similarly, repelling OECSs are tangent to the  eigenvector $\bm{e}_{-}$, and emanate from maxima of $s_{+}$, which quantify exactly their instantaneous repulsion rate. 
Equation \eqref{s1_ridge_criterion} implies that attracting iLESs are also parallel to $\bm{e}_{+}$ almost everywhere, while \eqref{s1_ridge_concavity_criterion} requires that the $\bm{e}_{+}$-line is within a concave trench of $s_{-}$. This further condition, therefore, can lead to cases in which there are attracting OECSs but not attracting iLESs. Similar considerations hold for repelling structures.
However, it should be noted that iLESs are not restricted to two-dimensional flows, as  OECSs are, 
but generalize to $n$-dimensional systems.

\section{Examples}\label{sec:examples}

In this section, several examples of the calculation of the iLE field (and iLESs) are given, with comparisons to the usual FTLE approach.
Section \ref{secnonlinearsaddle} considers a nonlinear saddle flow which can be worked out entirely analytically. 
Section \ref{sec:2ddg} considers the time-dependent double-gyre, for which the velocity field can be written analytically. Sections \ref{sec:wrf} and \ref{sec:mseas} examine iLESs in realistic time-dependent two-dimensional geophysical flows. 
Finally section \ref{sec:3dabc} explores the use of iLESs in a fully coupled three-dimensional flow.

\subsection{Two-Dimensional Nonlinear Saddle Flow}\label{secnonlinearsaddle}

\begin{figure}[ht]
    \centering
    \includegraphics[width=\textwidth]{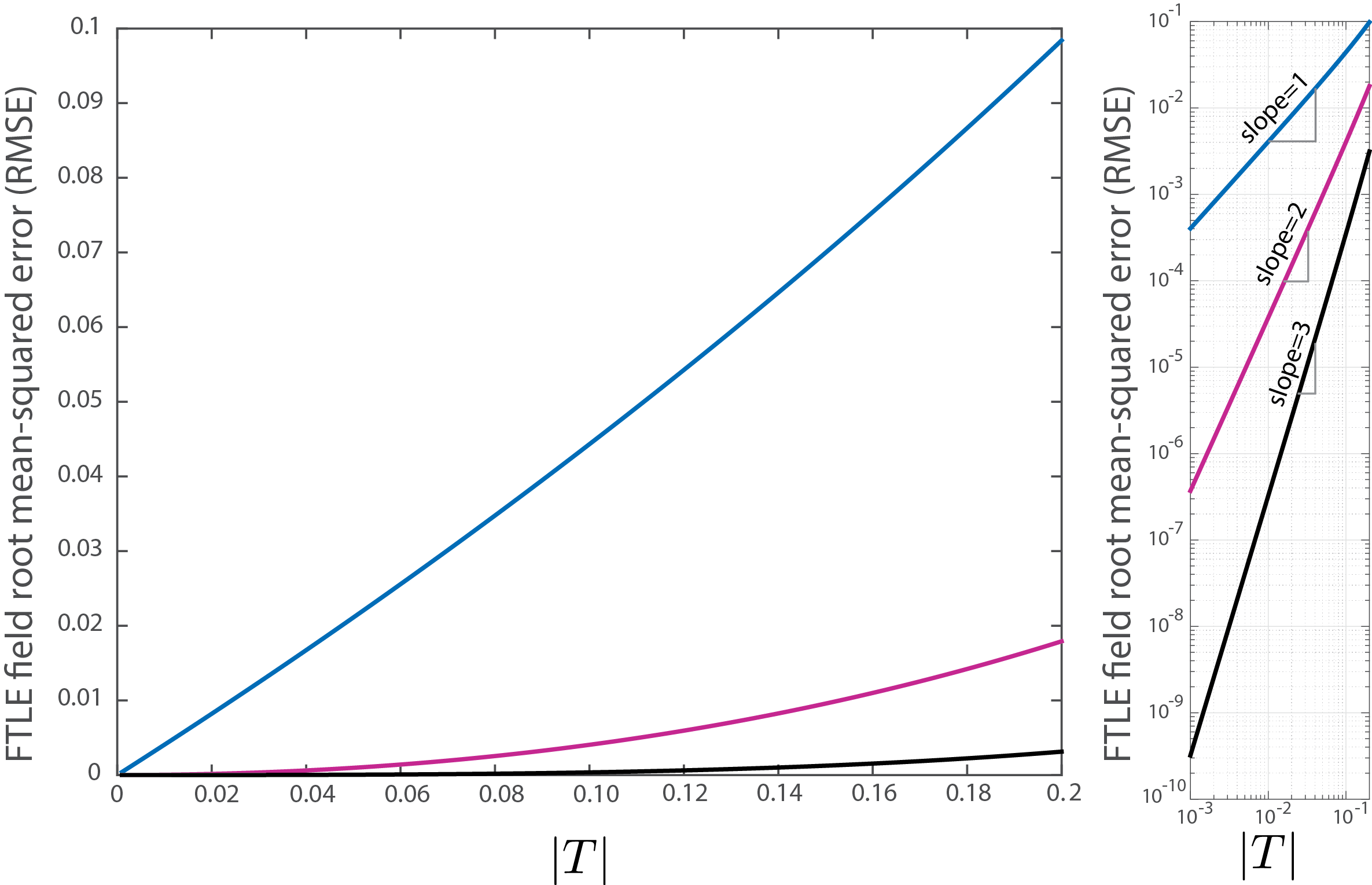}
    \caption{Root mean-squared error (RMSE) for successive approximations of the backward-time FTLE field for the nonlinear saddle \eqref{nonlinear_saddle} expanded in $T$: zeroth-order (blue), first-order (magenta), second-order (black). Notice that the error grows linear in $|T|$, quadratic in $|T|$, and cubic in $|T|$, respectively, as shown more clearly in the log-log plot on the right.}
    \label{fig:rmse_thru_order2_normal_and_loglog}
\end{figure}

Consider the following nonlinear saddle flow with cubic term,
\begin{equation}
\begin{split}
\dot x &= x, \\
\dot y &= -y + y^3,
\label{nonlinear_saddle}
\end{split}
\end{equation}
in the domain $U=\{(x,y) \in \mathbbm{R}^2~\big| ~ |y|<1 \}$.  A portion of the vector field is depicted in the inset of Fig.\ \ref{fig:rmse_thru_order2_normal_and_loglog}.
These two uncoupled ordinary differential equations admit the explicit solutions,
\begin{equation}
\begin{split}
x(t) &= x_0 e^t, \\
y(t) &= \frac{y_0}{\sqrt{(1 - y_0^2)e^{2t} + y_0^2}},
\label{nonlinear_saddle_solution}
\end{split}
\end{equation}
where the initial condition at time $t_0=0$ is $\mathbf{x}_0=(x_0,y_0)$.  
The right Cauchy-Green deformation tensor for a  backward integration time $T<0$, is,
\begin{equation}
\mathbf{C}_{0}^{T}(\mathbf{x}_0) =
\begin{bmatrix}
     e^{2T} & 0 \\ 
     0 & \frac{e^{4T}}{( (1 - y_0^2)e^{2T} + y_0^2)^3}  
\end{bmatrix},
     \label{nonlinear_saddle_C}
\end{equation}
which yields a backward time FTLE of
\begin{equation}
    \sigma_0^T(\mathbf{x}_0) = - \frac{1}{2T} \log \Bigg( \frac{e^{4T}}{( (1 - y_0^2)e^{2T} + y_0^2)^3} \Bigg).
    \label{nonlinear_saddle_FTLE}
\end{equation}
Using Taylor series approximations for small $|T|$, the backward FTLE can be written as an expansion in $T$ for  small $|T|$,
\begin{equation}
    \sigma_0^T(\mathbf{x}_0) = (1 - 3y_0^2) + 3  y_0^2 (1 - y_0^2) T - 2y_0^2(1 - y_0^2)(1 - 2y_0^2)T^2 + \mathcal{O}(T^3).
    \label{nonlinear_saddle_FTLE_approximation_main}
\end{equation}
See Appendix \ref{nonlinear_saddle_details} for details.

The FTLE can be approximated by the first, second, and third terms (the zeroth-order, first-order, and second-order in $T$, respectively) using the procedure outlined in section \ref{eigS_T}.
The key symmetric matrices in the expansion of $\mathbf{C}$ are $\mathbf{S}$, $\mathbf{B}$, and $\mathbf{Q}$, which are given explicitly by,
\begin{equation}
\begin{split}
\mathbf{S}(\mathbf{x}_0) &=
\begin{bmatrix}
     1 & 0 \\ 
     0 & (- 1 + 3y_0^2)   
\end{bmatrix},\\
\mathbf{B}(\mathbf{x}_0) &=
\begin{bmatrix}
     2 & 0 \\ 
     0 & (2 - 18y_0^2 + 24y_0^4)  
\end{bmatrix},\\
\mathbf{Q}(\mathbf{x}_0) &=
\begin{bmatrix}
     \tfrac{8}{3} & 0 \\ 
     0 &  (-\tfrac{8}{3} + 56y_0^2 - 192 y_0^4 + 160 y_0^6)   
\end{bmatrix},
     \label{nonlinear_saddle_SBQ}
\end{split}
\end{equation}
Note that $\mathbf{S}(\mathbf{x}_0)$ has  a minimum eigenvalue $s_{-}(x_0,y_0) = - 1 + 3y_0^2$, the negative of which matches the first  term of \eqref{nonlinear_saddle_FTLE_approximation_main}, as prescribed by \eqref{FTLE_T_correction2}. The eigenvalue $s_{-}$ has a corresponding normalized eigenvector $\bm{e}_{-}=[0,1]^{\top}$. As shown in Appendix \ref{nonlinear_saddle_details}, the formulas of section \ref{eigS_T} for approximating the true FTLE, \eqref{nonlinear_saddle_FTLE_approximation_main}, of the nonlinear saddle, \eqref{nonlinear_saddle}, through second-order in the integration time $T$ can be analytically verified for this example.

\begin{figure}[!t]
    \centering
    \includegraphics[width=0.8\textwidth]{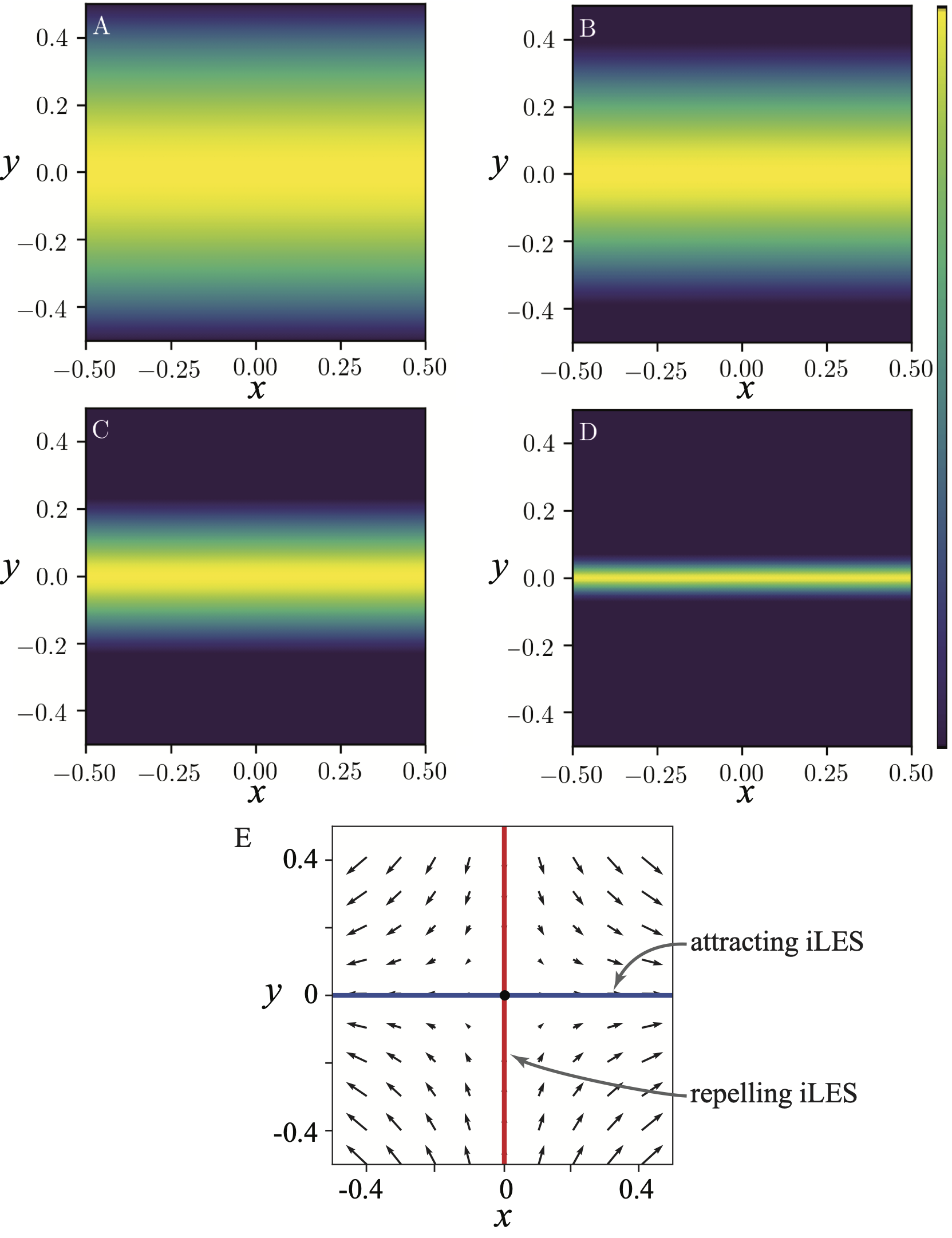}
    \caption{Comparison of the instantaneous attraction rate (A), with FTLE fields of backward non-dimensional integration times $T=$ (B) $-1$, (C) $-2$, and (D) $-4$. As the integration time magnitude increases increases, the average FTLE values decreases, thus  comparing the exact values of the heat-map is less meaningful than comparing the topography. For a topographical analysis, relatively high values are show in yellow and relatively low values in dark blue. A relative scale color bar is show on the right.
    (E)  The attracting and repelling iLESs are shown in blue and red, respectively, with the vector field overlaid.}
    \label{fig:nls_FTLE}
\end{figure}

To illustrate the accuracy of the successive approximations, Fig.\ \ref{fig:rmse_thru_order2_normal_and_loglog} shows the root mean-squared error (RMSE) for the FTLE field as a function of integration time magnitude, $|T|$, over the domain $U$.
Notice that, as expected, the error grows linear in $|T|$, quadratic in $|T|$, and cubic in $|T|$, for the zeroth-order, first-order, and second-order approximations, respectively.
The attracting rate and the backward time FTLE field for various integration times $T$ are shown in Fig.\ \ref{fig:nls_FTLE}.

Furthermore, we can demonstrate that the $x$-axis, $\{y=0\}$, is an attracting S-ridge. Note that  $s_{-}(x_0,y_0) = - 1 + 3y_0^2 < 0$ for $y_0=0$, satisfying the attraction criterion, \eqref{s1_criterion}. Since $\nabla s_{-}(x_0,y_0)=[0, 6y_0]^{\top}$, then along the $x$-axis we have $\nabla s_{-}(x_0,y_0)|_{y_0=0} \cdot \bm{e}_- = 0$, satisfying the ridge criterion, \eqref{s1_ridge_criterion}.  Furthermore, along the $x$-axis,
$(\textbf{H}_{s_{-}} \cdot \bm{e}_{-}) \cdot \bm{e}_{-}=6>0$, satisfying the concavity criterion, \eqref{s1_ridge_concavity_criterion}.  
The attracting and repelling iLESs are shown in Fig.\ \ref{fig:nls_FTLE}(E). 
 
\subsection{Two-Dimensional Time-Varying Double-Gyre Flow}\label{sec:2ddg}

While the time-varying  double-gyre does not admit an explicit solution, as the previous example does, one can still analytically approximate the FTLE field up to first-order in $T$ using the formulas of Section \ref{eigS_T}.

Consider the double-gyre flow as described in \cite{shadden2005definition}. This flow comes from the Hamiltonian stream function,
\begin{equation}
    \psi(x,y,t)= A\sin(\pi f(x,t))\sin(\pi y), 
\end{equation}
where,
\begin{equation}
f(x,t)=\epsilon\sin(\omega t)x^{2}+(1-2\epsilon\sin(\omega t))x. \label{f_function}
\end{equation}

The velocity field, $\mathbf{v}=(u,v)$, can be calculate as,
\begin{equation}
\begin{split}
    \dot x &= u(x,y,t) =   -\frac{\partial\psi}{\partial y} =   -A\pi\sin(\pi f(x,t))\cos(\pi y),\\
    \dot y &= v(x,y,t) = ~~ \frac{\partial\psi}{\partial x} = ~~ A\pi\cos(\pi f(x,t))\sin(\pi y)\frac{\partial f}{\partial x}(x,t).
    \label{doublegyre}
\end{split}
\end{equation}
The domain for $(x,y)$ is  $U=[0, 2]\times[0,1]$.
Following \cite{shadden2005definition}, parameters $A=0.1$, $\omega=0.2\pi$, and $\epsilon=0.25$ were chosen.  

From the gradient of this field (see Appendix \ref{doublegyre_details}), it can be analytically calculated via \eqref{s_analytical_explicit_fluid_incompressible} that the zeroth order approximation to the backward-time FTLE for an initial condition $\mathbf{x}_0=(x_0,y_0)$ at initial time $t_0$ in the infinitesimal integration time limit is,
\begin{equation}
\begin{split}
s_-&(\mathbf{x}_0,t_0) 
=
-\pi^2 A   \Bigg[  
 \epsilon^2 \sin ^2(\omega t) \Bigg\{
\sin^2(\pi y_0) \Big(  \sin^2(\pi f) (1-x_0)^2 
\\
& ~~ + \tfrac{1}{\pi}\sin(2\pi f) (1-x_0) +\tfrac{1}{\pi^2}\cos^2(\pi f)
\Big)
+  \cos^2(\pi y_0) \cos^2(\pi f) (1-x_0)^2 \Bigg\}
\\
& ~~  + \cos^2(\pi y_0) \cos^2(\pi f)  \Big(
1 - 4 \epsilon \sin (\omega t) (1-x_0) \Big)
~\Bigg]^{1/2}.
\end{split}
\label{doublegyre_s1_again}
\end{equation}
where the dependence of 
$f$ on $(\mathbf{x}_0,t_0)$ is understood. 
Note that the $s_-(\mathbf{x}_0,t_0)$ field, just like the vector field, is a periodic function of $t_0$ with period $2\pi/\omega$.  Note that for $t_0= k2\pi/\omega$, for some integer $k$, we have,
\begin{equation}
s_-(\mathbf{x}_0,t_0) = -\pi^2 A |\cos(\pi x_0) \cos(\pi y_0)|. \label{s1_doublegyre_t0_zero}
\end{equation}

\begin{figure}[!t]
\begin{center}    
\begin{tabular}{c@{\hspace{0.5pc}}c}
    \includegraphics[width=0.71\linewidth]{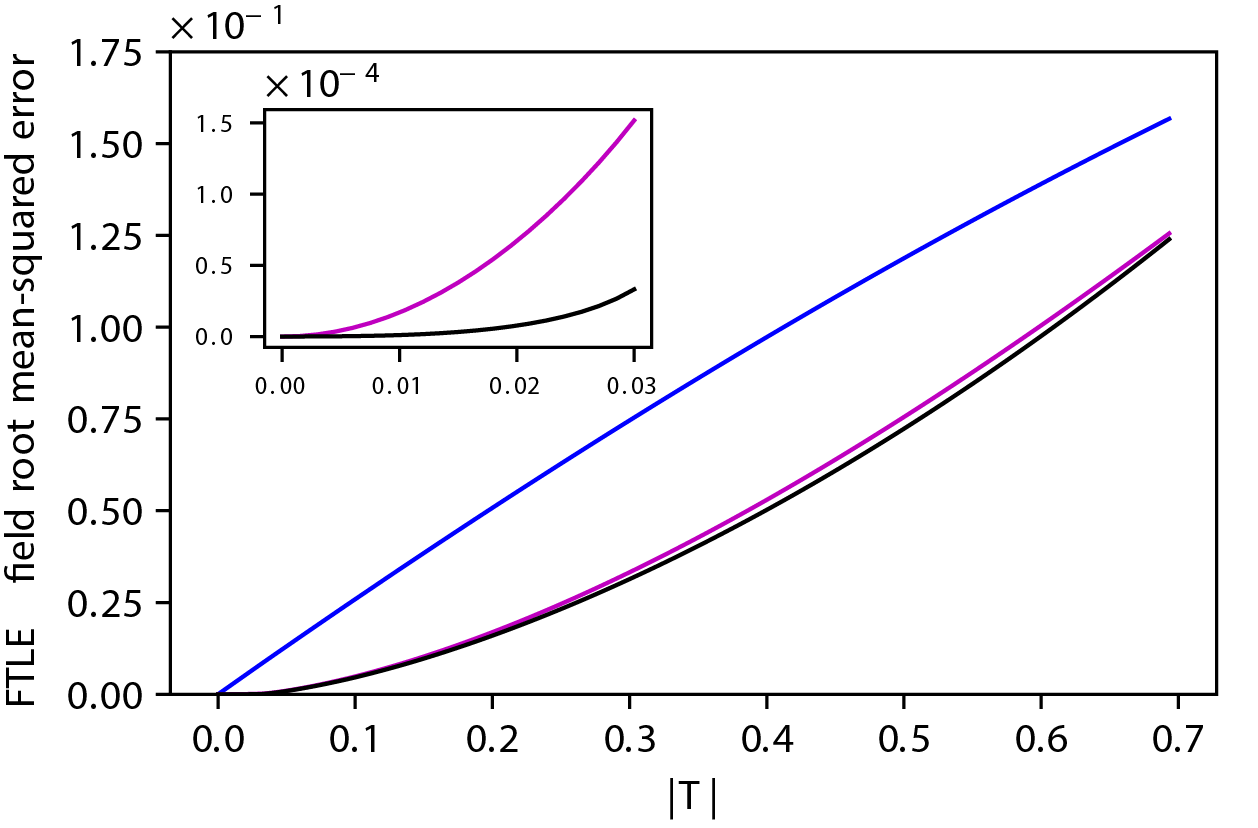}
     &
    \includegraphics[height=0.5\linewidth]{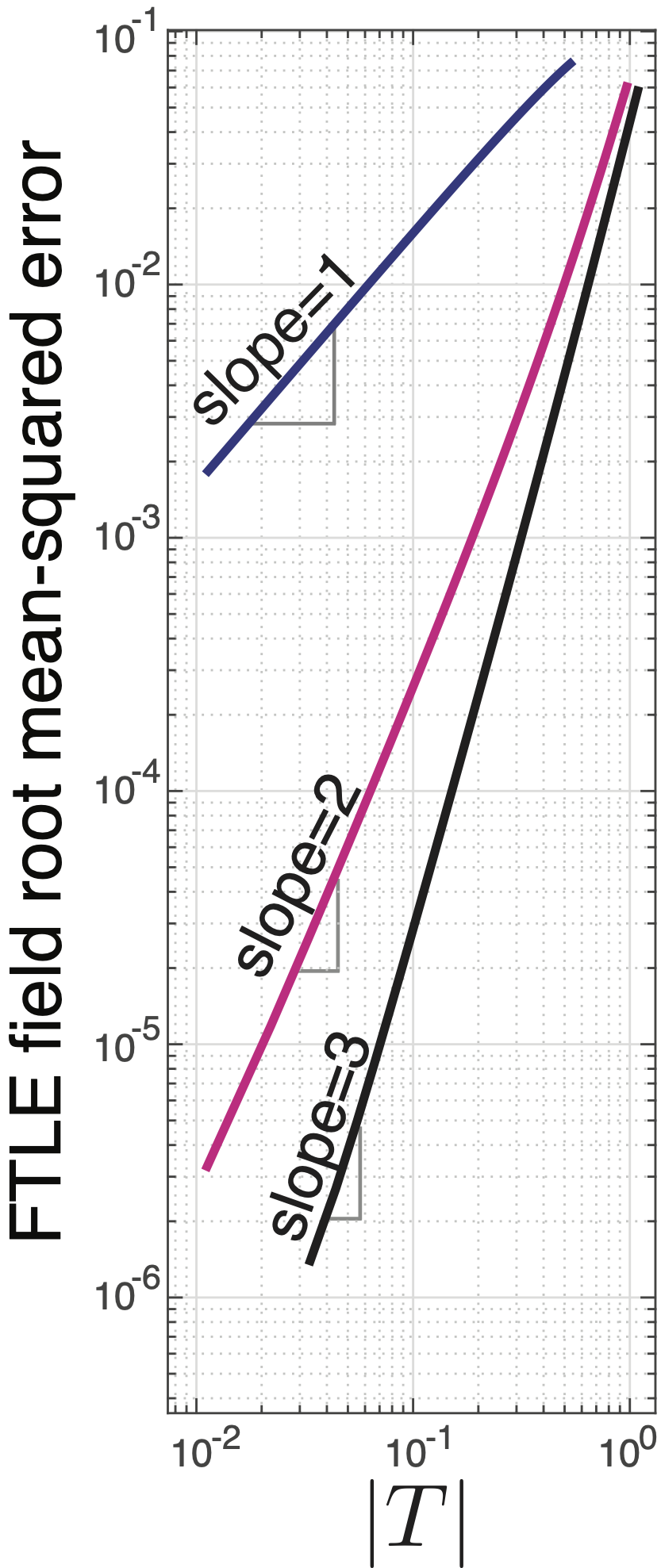}
    \end{tabular}
    \caption{Root mean-squared error (RMSE) vs.\ $|T|$ for successive approximations of the backward-time FTLE field for the double-gyre flow expanded in $T$: zeroth-order (blue), first-order (magenta), second-order (black), showing the error growing linear, quadratic, and cubic in $|T|$, respectively, as revealed more clearly in a log-log plot (right).}
    \label{fig:dg_rmse_ts}
    \end{center}
\end{figure}

The first-order term in the backward integration time $T<0$ can also be analytically determined. See Appendix \ref{doublegyre_details} for details.

\begin{figure}[!b]
 \centering
 \includegraphics[width=\linewidth]{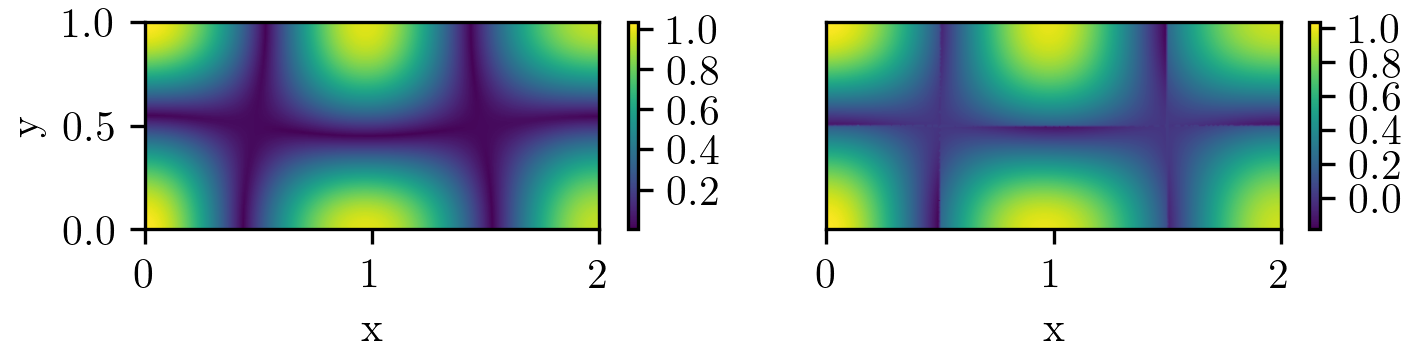}\\
 \includegraphics[height=0.24\linewidth]{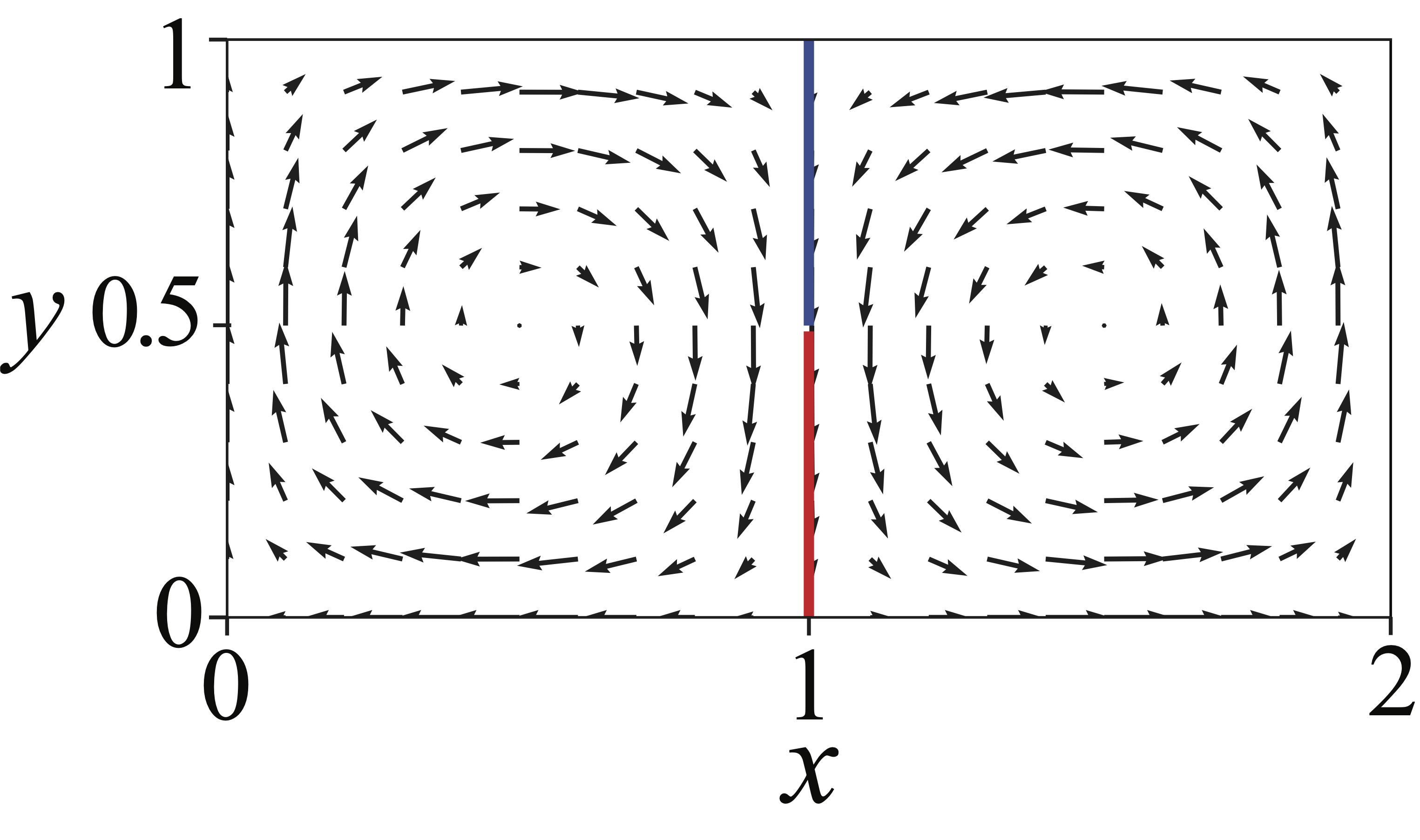}
 \caption{Top left: True backward-time FTLE field for the double-gyre  flow for an integration period of $T=-0.3$. Top right: The iLE approximation of the FTLE field to first-order in $T$, for the same integration time $T=-0.3$.  The root mean-squared error between these two fields is approximately 0.03 (see Fig.\ \ref{fig:dg_rmse_ts}). 
Bottom: The instantaneous Lyapunov exponent structures (iLESs) are shown---the attracting structure in blue and the repelling structure in red, both along $x=1$---with the vector field overlaid.
 Parameters: $A=0.1$, $\omega=0.2\pi$, $\epsilon=0.25$, and $t_0=0$.}
 \label{fig:dg_comp}
\end{figure}

Fig.\ \ref{fig:dg_rmse_ts} shows the root-mean-square error between the backward-time FTLE field and the zeroth order (blue), first order (magenta), and second order (black) approximations. 
In this figure, one can see that as the integration time, $|T|$, goes to 0, the approximations converge to the true (benchmark) FTLE field, as  expected. Note that the second-order term is more sensitive to numerical errors than either the zeroth- or first-order terms. Fig.\ \ref{fig:dg_comp} shows a comparison of the FTLE field for a short integration time, $T=-0.3$ (left), with an approximation to first order in $T$ (right).

As an incompressible two-dimensional flow, the infinitesimal time limit of the attraction and repulsion fields have the same structure (see \eqref{s_analytical_explicit_fluid}), differing only by a minus sign, $s_-(\mathbf{x}_0,t_0)=-s_+(\mathbf{x}_0,t_0)$.  However, because they have different eigenvector fields, their S-ridges according to the criteria in section \ref{sec:ilcs} are different.  For instance, at initial time $t_0=0$, $s_-(\mathbf{x}_0,0)$ and $-s_+(\mathbf{x}_0,0)$ are both given by the right-hand side of \eqref{s1_doublegyre_t0_zero}.  However, only the segment,
\begin{equation}
    \ell_a = \{ (x_0,y_0) \in U ~|~ x_0=1, 0.5<y_0<1 \},
\end{equation}
meets the criteria for an attracting iLES (see the blue curve in Fig.\ \ref{fig:dg_comp}, bottom panel). 
While criterion \eqref{s1_criterion} is met almost everywhere, along the line $x_0=1$, the eigenvector field $\bm{e}_-(\mathbf{x}_0,0)$ switches from $[1,0]^{\top}$ for $0.5<y_0<1$ to $[0,1]^{\top}$ for $0<y_0<0.5$, which leads to the  concavity criterion, \eqref{s1_ridge_concavity_criterion}, only being satisfied along $\ell_a$.
The situation is reversed for $s_+(\mathbf{x}_0,0)$, for which the eigenvector field $\bm{e}_+(\mathbf{x}_0,0)$ switches from $[0,1]^{\top}$ for $0.5<y_0<1$ to $[1,0]^{\top}$ for $0<y_0<0.5$,  which leads to the  concavity criterion, \eqref{sn_ridge_concavity_criterion}, only being satisfied along the segment,
\begin{equation}
    \ell_r = \{ (x_0,y_0) \in U ~|~ x_0=1, 0<y_0<0.5 \}
\end{equation}
which meets the criteria for a repelling iLES (see the red curve in Fig.\ \ref{fig:dg_comp}, bottom panel).  See Appendix \ref{doublegyre_details} for further details on the $s_-$ and $\bm{e}_-$ calculations.

\begin{figure}[b]
    \centering
    \includegraphics[width=0.8\linewidth]{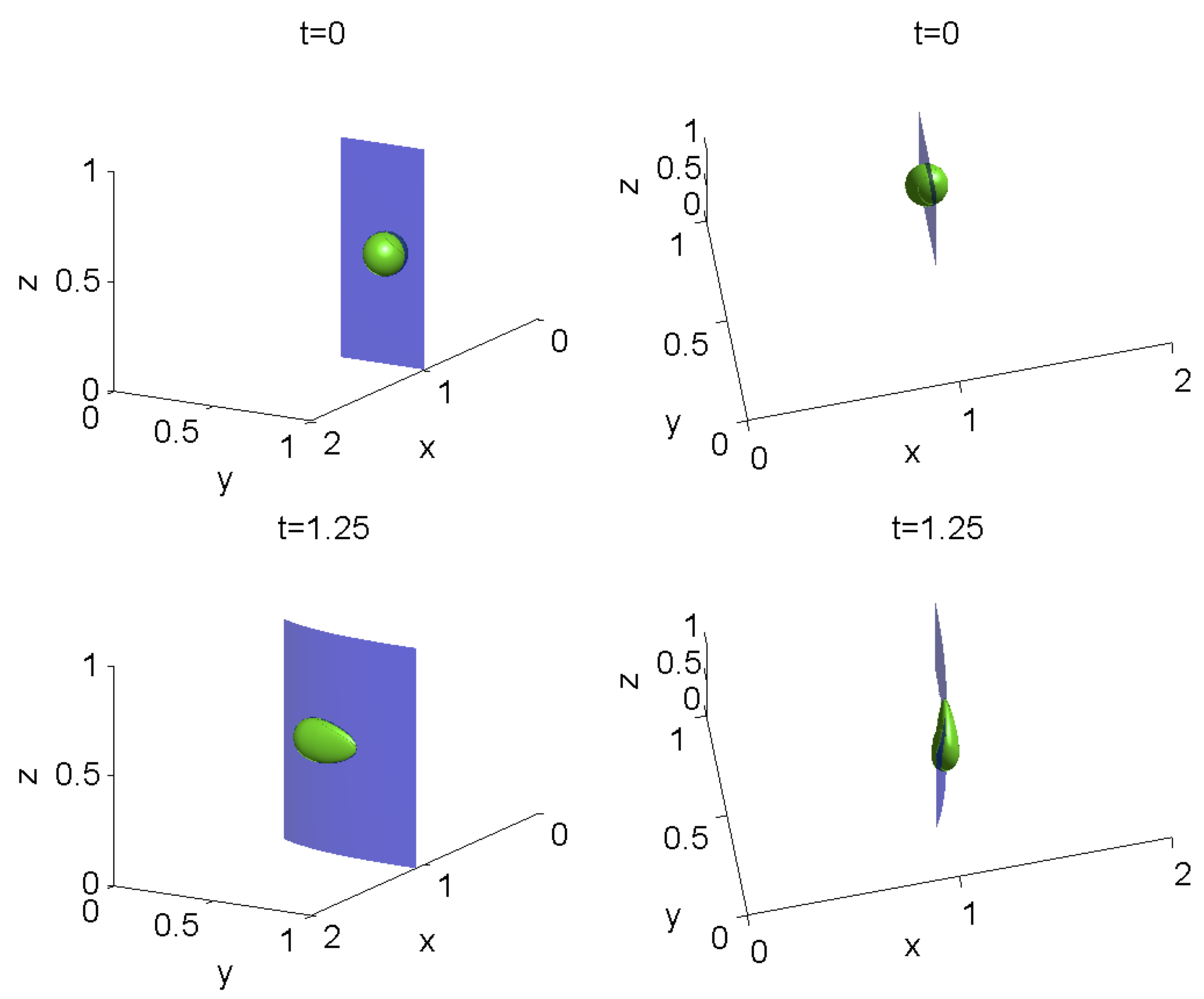}
    \caption{An attracting two-dimensional iLES surface, $\mathcal{S}_a$, blue, with a blob a passive tracers, green, shown from two different viewing angles. Top row shows the iLES and tracers at the initial time, $t_0=0$. Bottom row shows the iLES and tracers after being advected forward in time to  $t=1.25$.
    An animation for the  attracting iLES is at \url{https://youtu.be/NWxdG7BY0_o}.
    }
    \label{fig:dg3d_ailcs}
\end{figure}

\begin{figure}[t]
    \centering
    \includegraphics[width=0.8\linewidth]{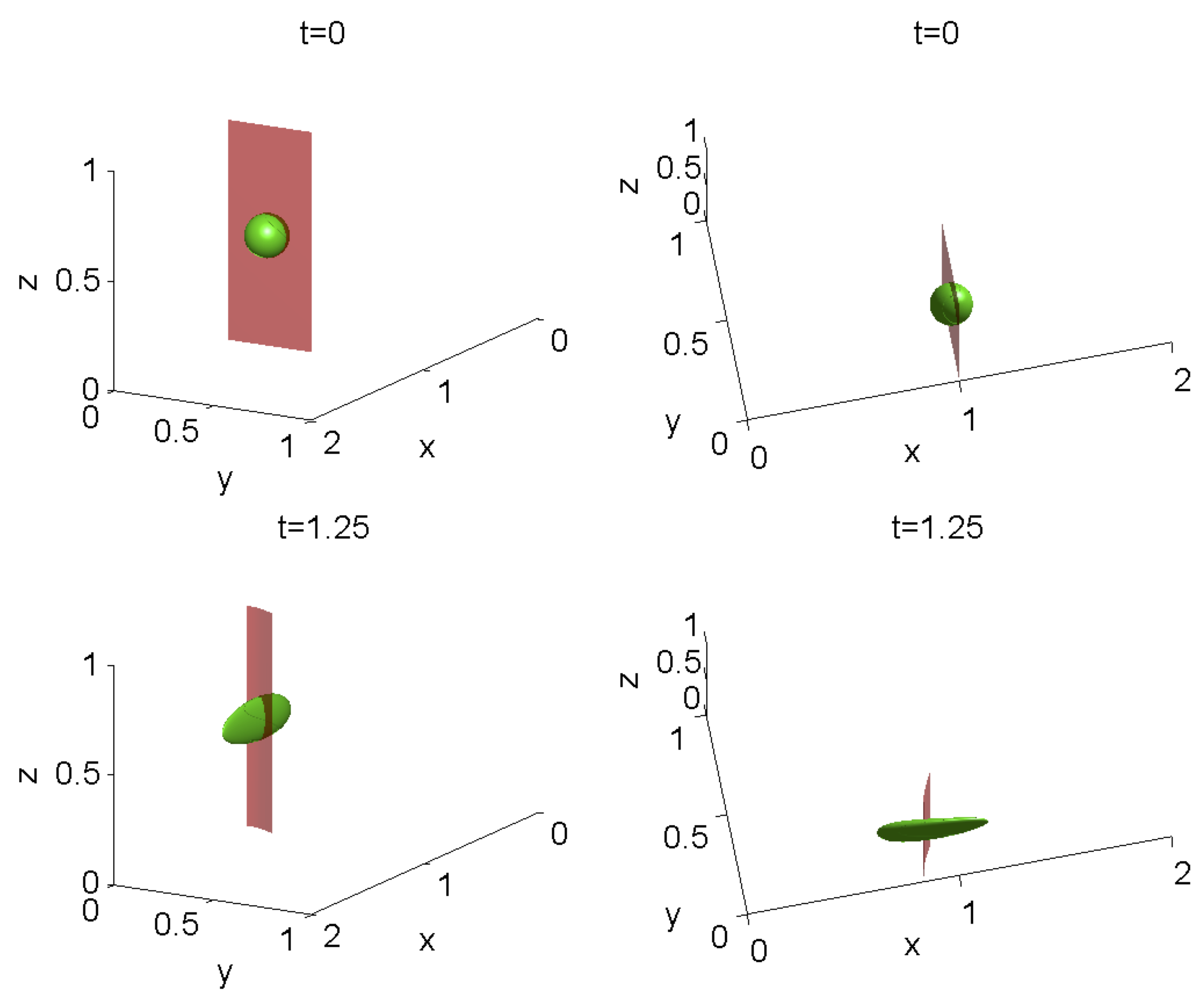}
    \caption{A repelling two-dimensional iLES surface, $\mathcal{S}_r$, red, with a blob a passive tracers, green, from two different viewing angles. Top row shows the iLES and tracers at the initial time, $t_0=0$. Bottom row shows the iLES and tracers after being advected forward in time to $t=1.25$.
    The animation for the repelling iLES is at \url{https://youtu.be/ZkD3qBnrHL0}.}
    \label{fig:dg3d_rilcs}
\end{figure}

These segments and their effect on nearby fluid particles are better visualized in an extension of the double-gyre flow to 3 dimensions, where we add a $z$ direction with no dynamics, $\dot z = 0$.
The attracting and repelling segments, $\ell_a$ and $\ell_r$, are extended to two-dimensional surfaces, $\mathcal{S}_a$ and $\mathcal{S}_r$ for the three-dimensional double-gyre flow and are shown in Figs.\ \ref{fig:dg3d_ailcs} and \ref{fig:dg3d_rilcs}, respectively.
Fig.\ \ref{fig:dg3d_ailcs} shows an attracting iLES (blue), along with a blob of passive tracers (green). Meanwhile, Fig.\ \ref{fig:dg3d_rilcs} show a repelling iLES (red), along with blob of tracers (green). In both figures, the first row shows the the initial configuration from two different angles, while the second row shows the configuration after being advected by the flow for a time of 1.25 in non-dimensional units. In Fig.\ \ref{fig:dg3d_ailcs}, one can see that the green blob, starting out as a sphere around a portion of the iLCS, becomes squeezed towards and spread along the iLES as the two are advected by the flow. In Fig.\ \ref{fig:dg3d_rilcs}, the green blob, starting as a sphere, spreads out and away from the repelling iLES as they are advected by the flow. These behaviors demonstrate that iLES are indeed the instantaneous approximation of traditional FTLE ridges in three dimensions. 
An animation for the  attracting iLES can be found at \url{https://youtu.be/NWxdG7BY0_o}, and the repelling iLES at \url{https://youtu.be/ZkD3qBnrHL0}.

\subsection{Two-Dimensional Oceanographic Flow Example}\label{sec:mseas}


In this section, a realistic oceanic flow model is employed to explore the methods described above, using ocean surface velocity data from a Global Real-Time Ocean Forecast System (Global RTOFS) \cite{RTOFS_web} model simulation for the Gulf of Mexico. This model was run with a horizontal grid resolution of $(\tfrac{1}{12})^{\circ}$  
and temporal resolution of 1 hour, which was then interpolated to a Cartesian grid with a horizontal resolution of 10 km. The fluid simulation forecast was for a $72$ hour period beginning at 0000 UTC 25 July 2019, i.e., $I=[$0000 UTC 25 July 2019, 0000 UTC 28 July 2019$]$.

\begin{figure}[!t]
    \centering
    \includegraphics[width=\textwidth]{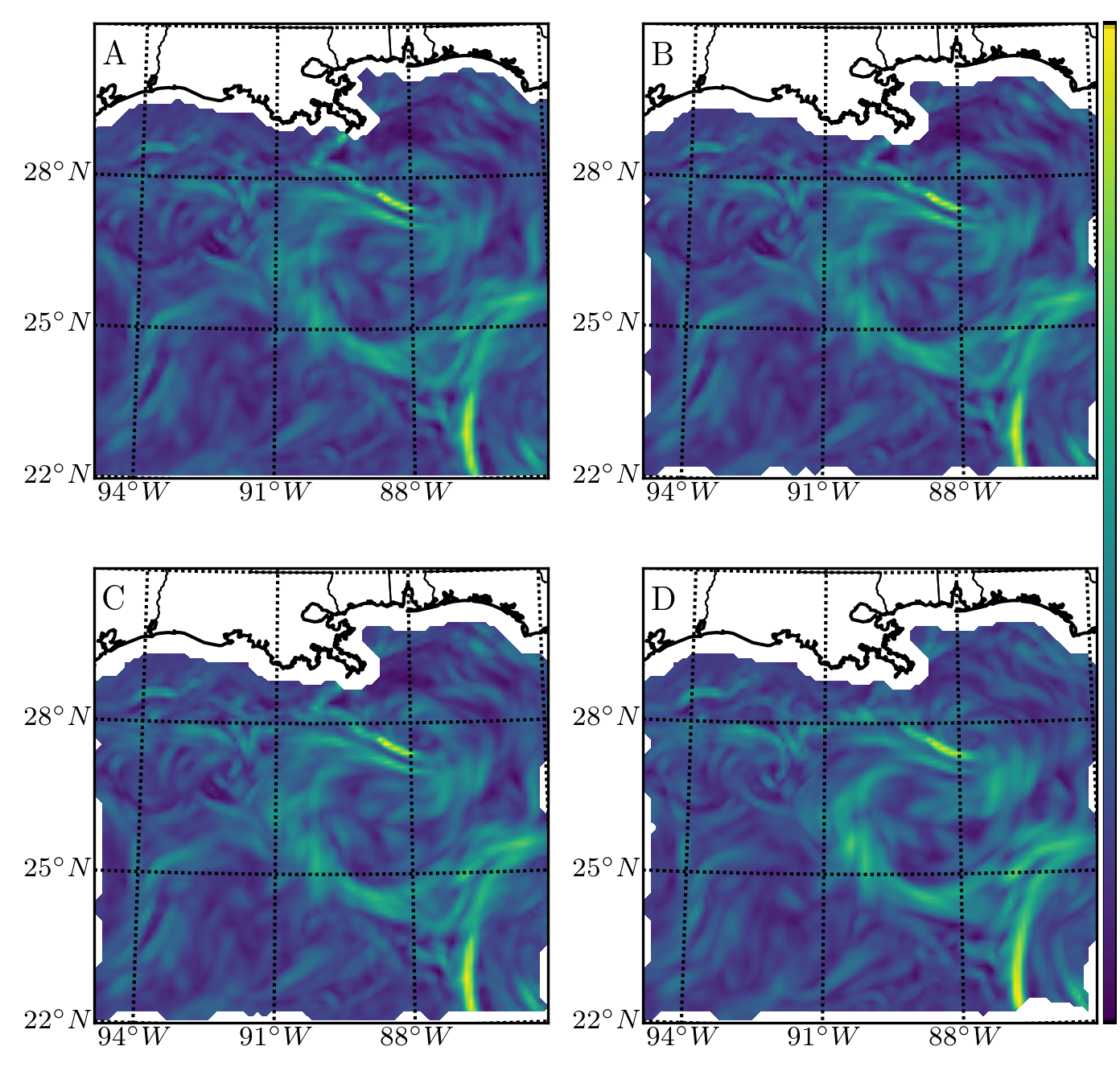}
    \caption{Comparison of the instantaneous attraction rate (A), with FTLE fields of integration times $T=$ (B) $-1$ hour (C) $-4$ hour, (D) $-24$ hour. As the integration time magnitude increases increases, the average FTLE values decreases, thus  comparing the exact values of the heat-map is less meaningful than comparing the topography. For a topographical analysis, relatively high values are show in yellow and relatively low values in dark blue. A relative scale color bar is show on the right. The spatial correlation between these fields is shown in Fig.\ \ref{fig:mv_corr}.}
    \label{fig:s1_Backward-Time_FTLE_Comparison_mv}
\end{figure}

\begin{figure}[!t]
    \centering
    \begin{tabular}{c@{\hspace{0.5pc}}c@{\hspace{0.5pc}}c}
    \includegraphics[width=0.395\textwidth]{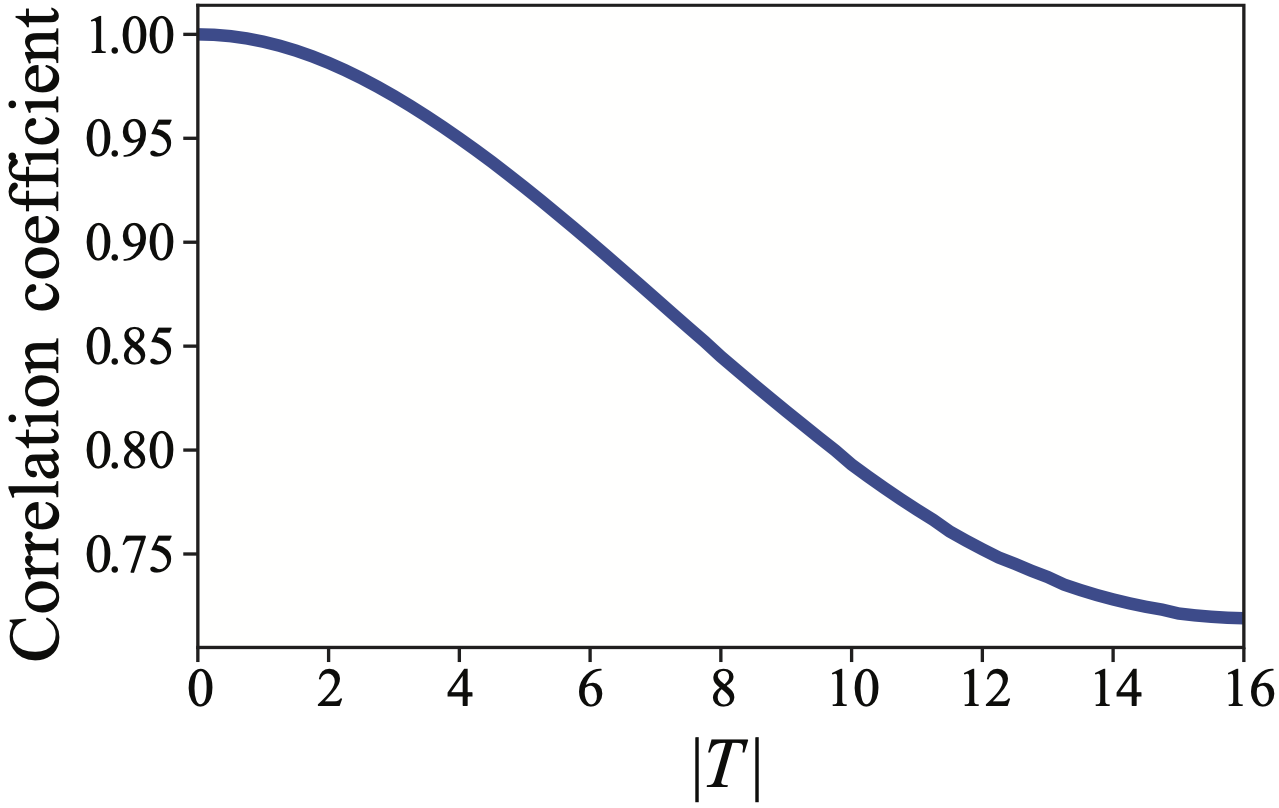} &
    \includegraphics[width=0.375\textwidth]{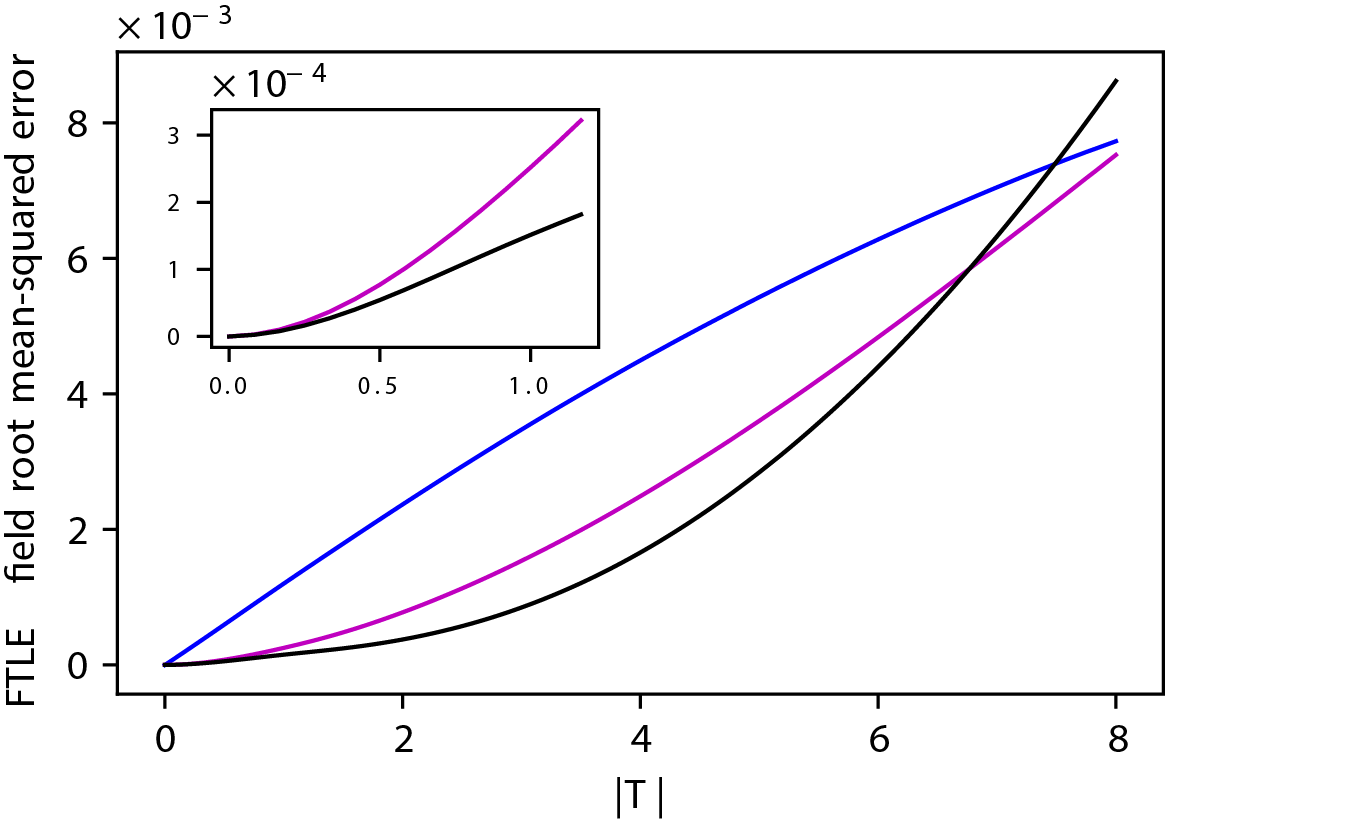} &
    \includegraphics[height=0.255\textwidth]{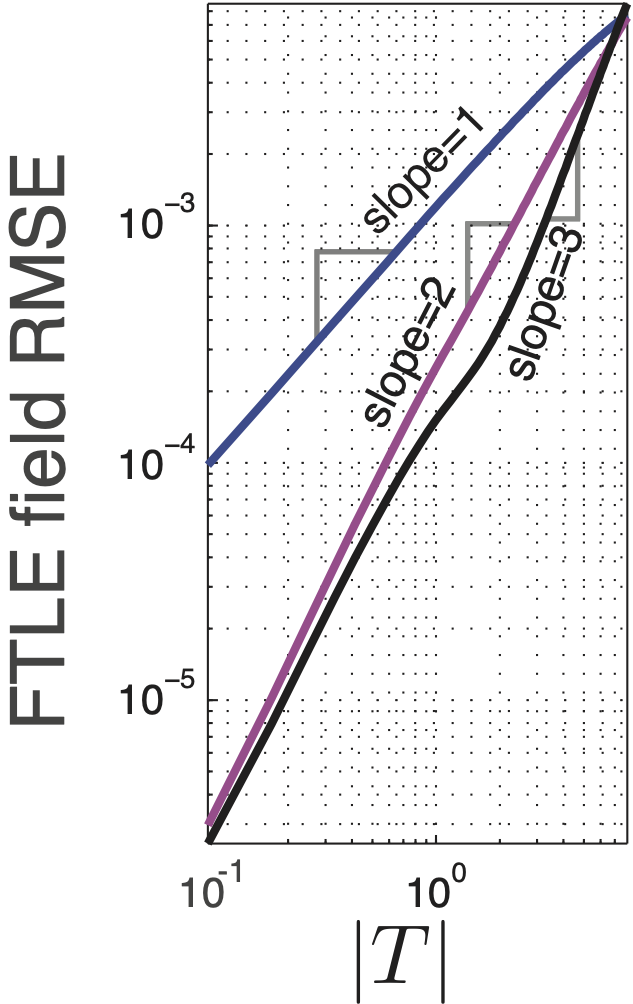}\\\\
    (A) & (B) & (C)
    \end{tabular}
    \caption{(A) Correlation coefficient between the attraction rate field and the benchmark backward-time FTLE field as a function of integration time, $|T|$, in hours.
    (B)
    RMSE for successive approximations of the backward-time FTLE field for an oceanic flow expanded in $T$: zeroth-order (blue), first-order (magenta), and second-order (black). Time is in hours. The inset shows the behavior for the higher order terms for $|T|$ close to $0$.
    (C) Same as (B), but on a log-log scale.}
    \label{fig:mv_corr}
\end{figure}

Fig.\ \ref{fig:s1_Backward-Time_FTLE_Comparison_mv} visually explores the connection between the attraction rate and the FTLE field in a two-dimensional oceanic flow. 
Panel A shows the attraction rate field at $t_0=$ 0000 UTC 26 July 2019. Panels B, C, and D show the FTLE field for 1, 4, and 24 hours of backward-time integration, initialized at time $t_0$. 
It can be seen that the significant Lagrangian transport structures over the time interval examined are already present in the attraction rate field. As the field is integrated backward in time the transport structures become sharper and grow longer, but do not change significantly. 
As the integration time is increased, the transport patterns which are shown by the attraction rate field become more sharply defined. This relationship can be quantified by the Pearson correlation coefficient, which is given in Fig.\ \ref{fig:mv_corr}(A). This figure shows that for integration times up to 16 hour, there is a strong correlation ($>0.7$) between the attraction rate and backward-time FTLE field. 

\begin{figure}[!t]
    \centering
    \includegraphics[width=\textwidth]{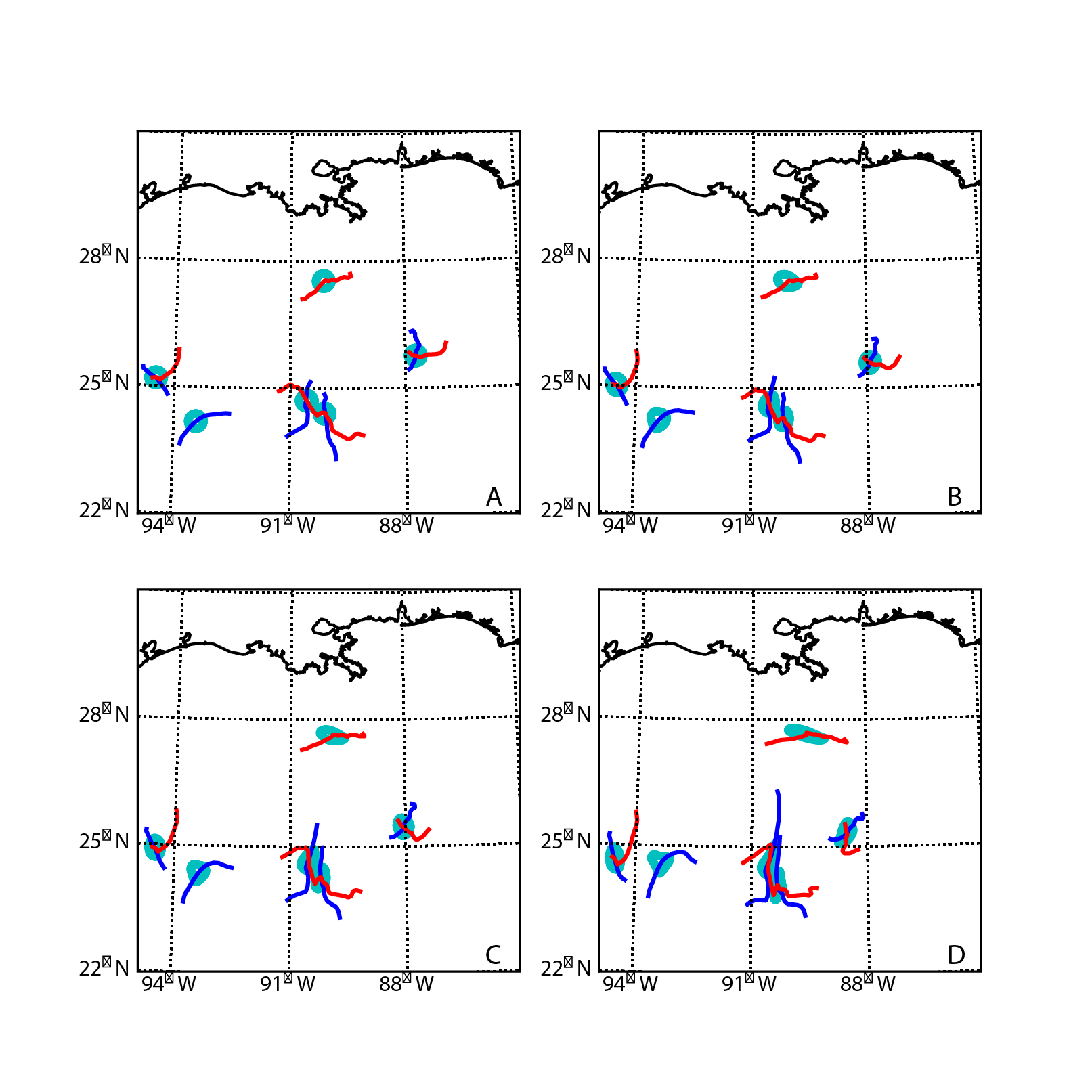}
    \caption{The iLESs  and passive tracers near the coast of Louisiana at different elapsed times, $T=t-t_0$. (A) $T=0$ hour, (B) $T=12$ hour, (C) $T=24$ hour, (D) $T=48$ hour. Repelling iLESs are shown in red, attracting iLES in blue, and passive tracers in cyan.}
    \label{fig:mv_iLCS}
\end{figure}

This data set was also able to numerically verify the relationship between the attraction rate, higher-order instantaneous approximations, and the backward-time FTLE field for a two-dimensional oceanic fluid flow. This result is presented in Fig.\ \ref{fig:mv_corr}(B) and (C), which shows the RMSE of these approximations compared with a benchmark FTLE field, 
where integration is performed backward in time from $t_0=$ 0000 UTC 26 July 2019. 
The blue line shows the RMSE for the attraction rate, the magenta for the attraction rate with a first-order correction term, and the black for the attraction rate with a second-order correction. As $|T|$ goes to $0$, the RMSE of all three approximations also goes to $0$, thus numerically verifying the relationships in section \ref{eigS_T} applied to a two-dimensional oceanic flow.

For this flow, it was further possible to verify that iLESs are effective at predicting the Lagrangian behavior of passive tracer particles advected in a two-dimensional oceanic flow. Fig.\ \ref{fig:mv_iLCS} shows the evolution of attracting iLESs (blue), repelling iLESs (red), and some passive tracers (cyan). 
These structures were initialized at $t_0=$ 0000 UTC 25 July 2019 (24 hours before the initial time given in Fig.\ \ref{fig:s1_Backward-Time_FTLE_Comparison_mv}) and advected forward in time. 
The structures present in the flow depend on the initial time $t_0$.
This different initial time was chosen for illustrative purposes as it shows a larger variety of structures present in the flow.
Panel A shows the iLESs and tracers at the initial time. Panels B, C, and D show the iLESs and tracers after $12$, $24$, and $48$ hours, respectively. In these panels, one can see that as time moves forward passive tracers are repelled away from the repelling iLESs and attracted towards the attracting iLESs, as is expected.

\subsection{Two-Dimensional Atmospheric Flow Example}\label{sec:wrf}


In this section, the methods described above are applied to a realistic time-varying atmospheric flow example, using wind data from a Weather Research and Forecasting (WRF) model simulation over the southeastern United States \cite{wrf_web}. This model was run with a horizontal grid resolution of 12 km 
and temporal resolution of 1 hour. Due to the scale mismatch between the horizontal resolution and the vertical resolution (which varies between 0.05 and 1 km), a single vertical level was chosen to focus on for this analysis. The level that was chosen corresponds to approximately 100 m above ground level (AGL), similar to what has been done in previous atmospheric LCS studies \cite{bozorgmagham2013real,tallapagada2011lagrangian,schmale2012isolates,bozorgmagham2015atmospheric}, as this is a level reachable by unmanned aerial vehicles for in situ meteorological measurements and sampling \cite{nolan2018coordinated,barbieri2019small}. The simulation data is available  for a $48$ hour interval 
$I = [$0000 UTC 30 June 2011, 0000 UTC 2 July 2011$]$.

\begin{figure}[!t]
    \centering
    \begin{tabular}{c@{\hspace{0.5pc}}c}
    \includegraphics[width=0.45\textwidth]{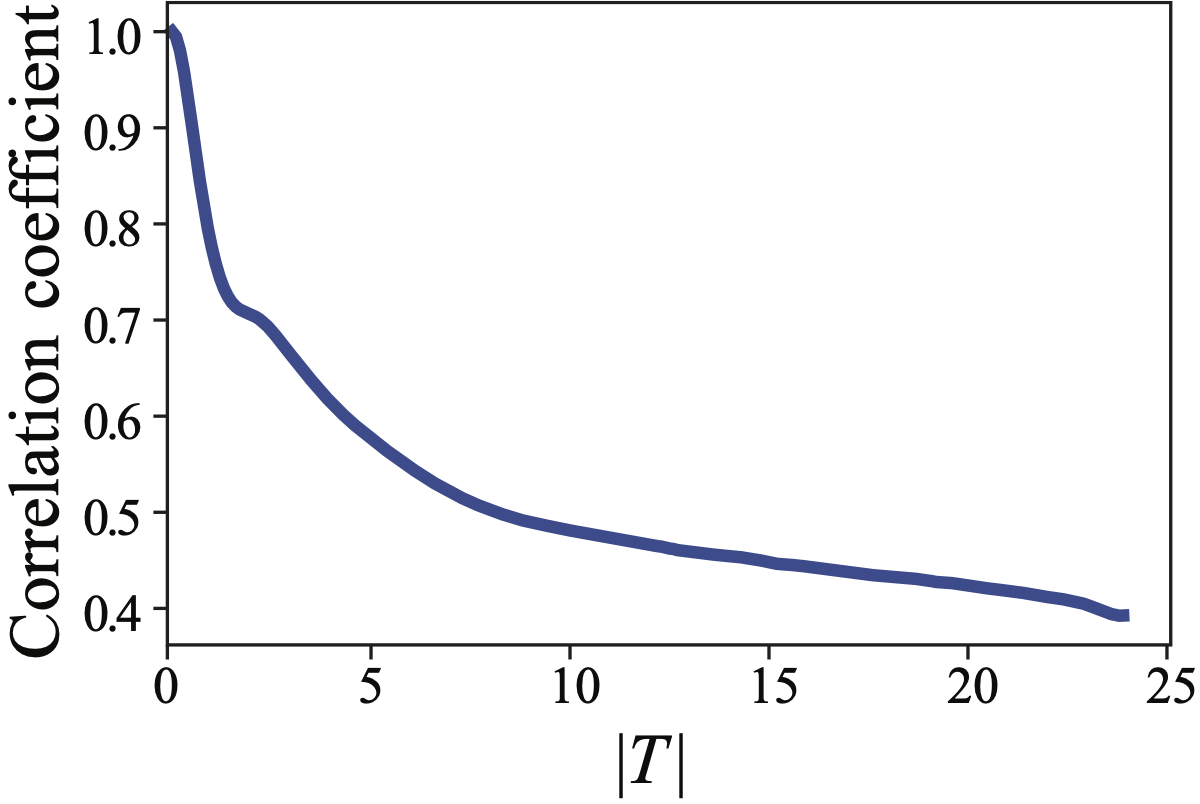} &
    \includegraphics[width=0.52\linewidth]{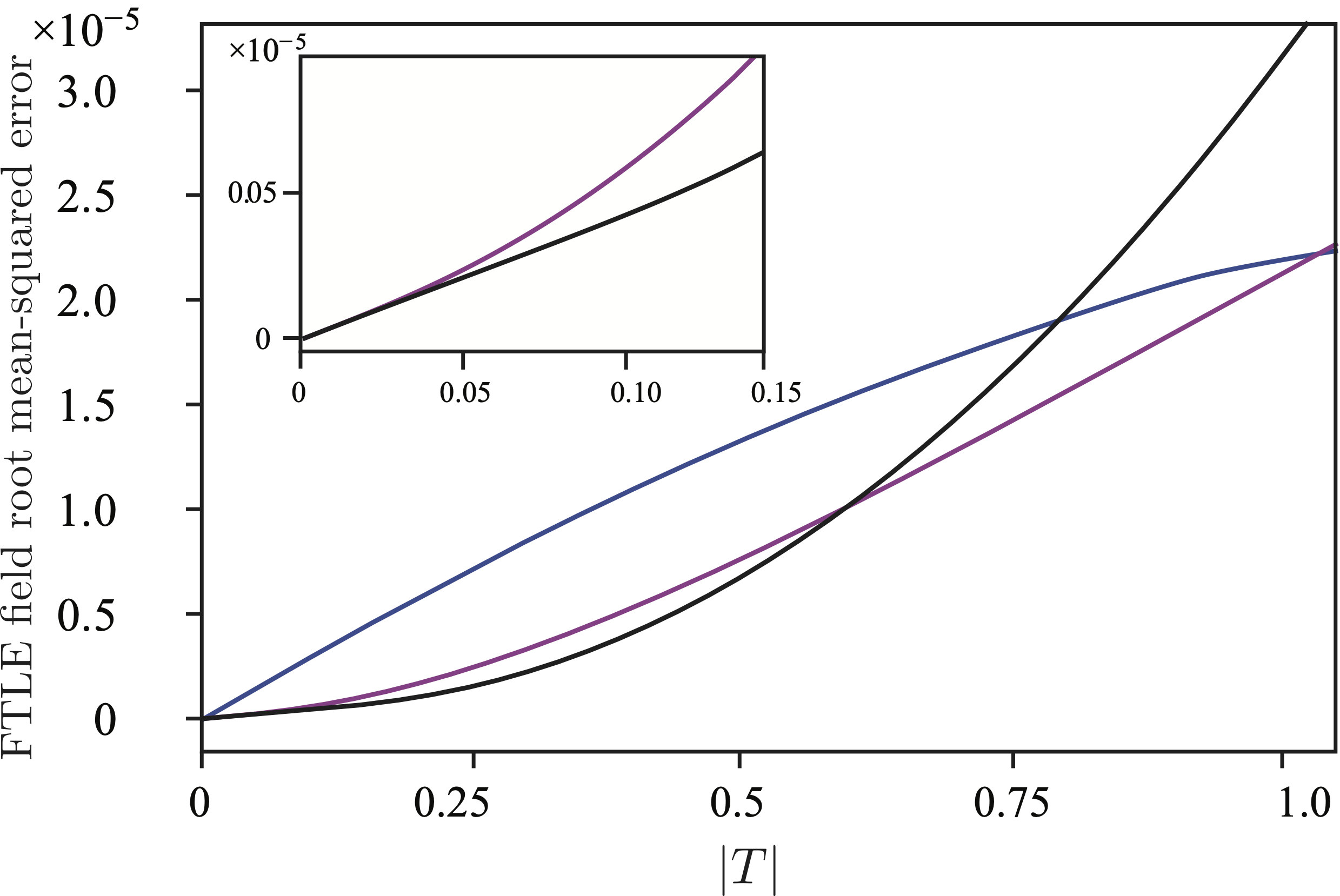}\\
    (A) & (B) 
    \end{tabular}
    \caption{(A) Pearson correlation coefficient between the attraction rate field and the benchmark backward-time FTLE field as a function of  integration time, $|T|$, in hours.
    (B) RMSE for successive approximations of the backward-time FTLE field for an atmospheric flow expanded in $T$: zeroth-order (blue), first-order (magenta), and second-order (black). Time is in seconds. The inset shows the behavior for the higher order terms for $|T|$ close to $0$.}
    \label{fig:wrf_corr}
\end{figure}

Using this data set the relationship between the attraction rate, higher-order  iLE approximations, and the particle-integration-based backward-time FTLE field for a two-dimensional atmospheric fluid flow can be numerically verified. This can be seen in Fig.\ \ref{fig:wrf_corr}(B), which shows the RMSE of these approximations with the FTLE field as integration is performed backward in time from an initial time $t_0$  in the interval $I$.
The blue line shows the RMSE for the attraction rate, the magenta for the attraction rate with the correction term to first-order in $T$, and the black for the attraction rate with the correction term to second-order in $T$. As $|T|$ goes to 0, the RMSE of all three approximations also goes to 0, thus numerically verifying the relationships shown in section \ref{eigS_T} for a two-dimensional atmospheric flow. This figure also shows that for small $|T|$, the second-order approximation is the most accurate, as expected. However, for larger $|T|$ the attraction rate will provide the most accurate approximation. 

\begin{figure}[t!]
    \centering
    \includegraphics[width=\textwidth]{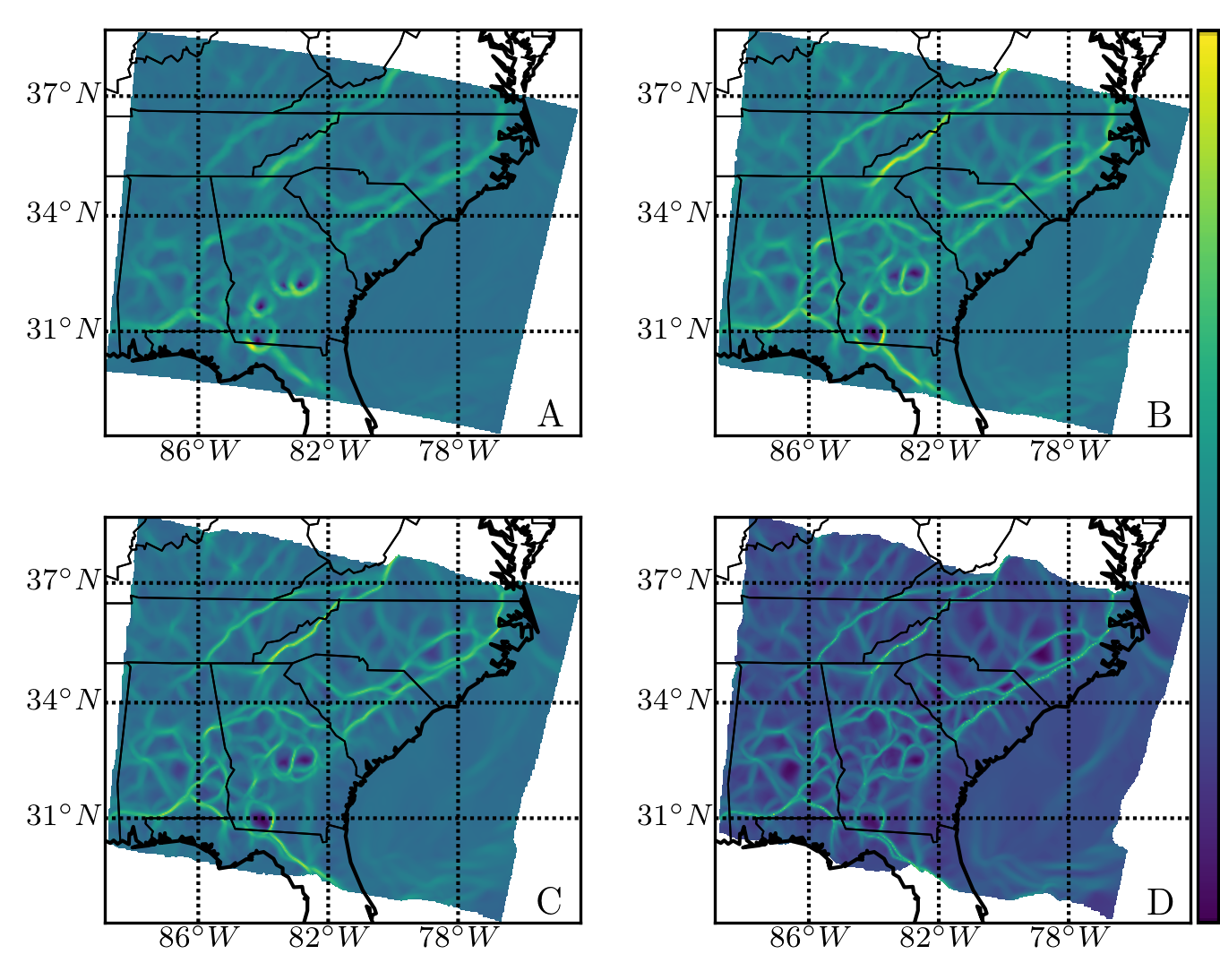}
    \caption{Comparison of the attraction rate (A), with FTLE fields of integration times $T=$ (B) $-1$ hour (C) $-2$ hour, (D) $-4$ hour. As the integration time increases increases the average FTLE values decreases, thus the comparing the values of the heat-map is less meaningful than comparing the topography. For a topographical analysis, relatively high FTLE values are show in yellow and relatively low values in dark blue (a relative scale color bar is show on the right). The spatial correlation between these fields  is given in Fig.\ \ref{fig:wrf_corr}.
    An animation of the comparison of the attraction rate with the (backward-time) attracting FTLE field of integration times up to 24 hours in backward time, in 10 minute increments, can be found at \url{https://youtu.be/nkqpZ2GcO1E}.}
    \label{fig:s1_Backward-Time_FTLE_Comparison}
\end{figure}

\begin{figure}[!t]
    \centering
    \includegraphics[width=\textwidth]{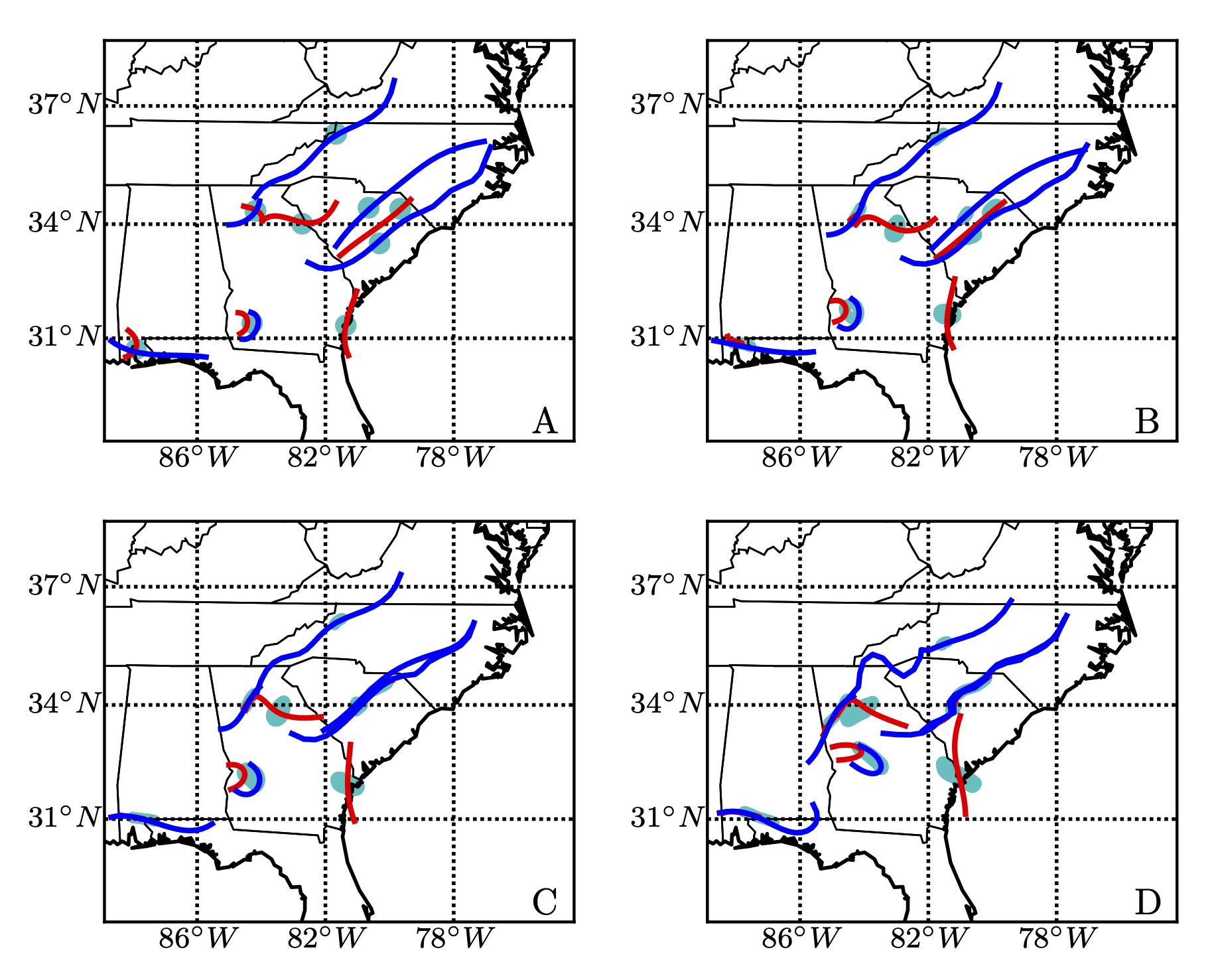}
    \caption{The iLESs and passive tracers at different elapsed times, $t$, since the initial evaluation time, $t_0=$ 0000 UTC 1 July 2011, of the simulation. (A) $t-t_0=0$ hour, (B) $t-t_0=2$ hour, (C) $t-t_0=4$ hour, (D) $t-t_0=8$ hour. Repelling iLESs are shown in red, attracting iLESs in blue, and passive tracers in cyan. An animation of the evolution of the iLESs and tracers over the entire 24 hour period can be found at \url{https://youtu.be/h4UhJT8vsiU}.
    }
    \label{fig:SE_iLCS}
\end{figure}

Fig.\ \ref{fig:s1_Backward-Time_FTLE_Comparison} visually explores the connection between the attraction rate and the FTLE field.
Panel A shows the attraction rate field at $t_0=$ 0000 UTC 1 July 2011. Panels B, C, and D show the FTLE field for 1, 2, and 4 hours of backward-time integration. In these plots, it can be seen that the important Lagrangian transport structures over the period examined are already present in the attraction rate field, even though this accuracy is not reflected in the RMSE plot, Fig.\ \ref{fig:wrf_corr}(B). As the field is integrated backward in time the transport structures become sharper and grow longer, but do not change significantly. For this particular flow, as the integration time is increased the transport patterns which are shown by the attraction rate field become more sharply defined. This relationship can be quantified by the Pearson correlation coefficient, given in Fig.\ \ref{fig:wrf_corr}(A), which shows that for short integration times ($<4$ hours), there is a strong correlation ($>0.6$) between the attraction rate and backward-time FTLE field. Then, as the integration time is increased the correlation between the fields becomes weaker. However, note that even for $|T|=$ 24 hours, there is still a moderate correlation ($>0.4$), not yet nearing zero.
An animation of the comparison of the attraction rate with the (backward-time) attracting FTLE field of integration times up to 24 hours in backward time can be found at \url{https://youtu.be/nkqpZ2GcO1E}.

This data set allows us to test whether iLESs are effective at predicting the Lagrangian behavior of passive tracer particles advected in a two-dimensional atmospheric flow, even though they are only evaluated at an initial time, $t_0=$  0000 UTC 1 July 2011. Fig.\ \ref{fig:SE_iLCS} shows the evolution of attracting iLESs (blue), repelling  iLESs (red), and some example passive tracers (cyan). 
These structures were initialized at $t_0$ and advected forward in time. Panel A shows the iLESs and tracers at the initial time. Panels B, C, and D show the iLESs and tracers after 2, 4, and 8 hours, respectively. An animation of the evolution of the iLESs and tracers over the entire 24 hour period can be found at \url{https://youtu.be/h4UhJT8vsiU}. In these panels it can be seen that as time marches forward passive tracers are repelled away from the repelling iLESs and attracted towards the attracting iLESs, as expected. 

\begin{figure}[!h]
    \centering
    \includegraphics[width=\textwidth]{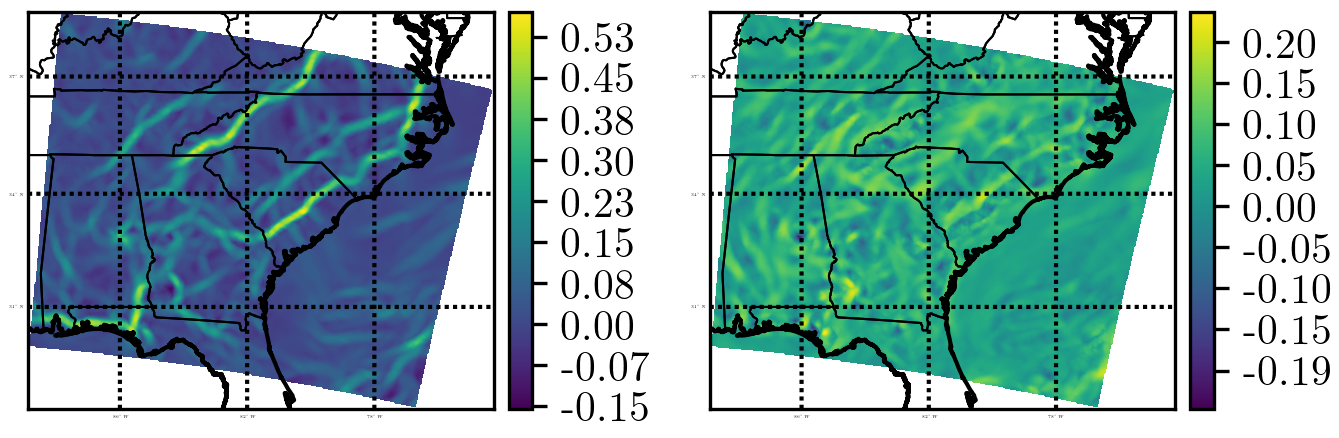}
    \caption{Comparison of the attraction rate field, $s_{1}$, left, and the repulsion rate field, $s_{2}$, right, at $T=0$. Structures in the attraction rate field are noticeably stronger than in the repulsion rate field. The attraction rate field has been multiplied by $-1$ to aid in visualization. The colorbar has units of hr$^{-1}$.}
    \label{fig:s1_vs_s2}
\end{figure}
Interestingly, it can also be seen that some of the repelling iLESs are attracted onto and effectively consumed by the attracting iLESs. A partial explanation for this can be found in Fig.\ \ref{fig:s1_vs_s2}, where a comparison between the attraction rate and the repulsion rate fields is shown.  
Note that the two-dimensional vector field on this level is not divergence-free, as the ignored vertical velocity is non-zero. Thus, the two fields are different (recall they would be the same if the vector  field was divergence-free, according to \eqref{s_analytical_explicit_fluid_incompressible}). In this figure, it can be seen that the attraction rate field is stronger than the repulsion rate field is; that is, the most attractive points of the attraction rate field are more than twice as attractive as the most repelling points in the repulsion rate field are repulsive.  Thus, it can be concluded that while the repelling iLESs are repulsive, the attracting iLESs are more attractive and thus overpower the repelling iLESs after a sufficient period of time. In similar applications, OECSs have successfully predicted short-term transport in several geophysical flows \cite{Serra2016,serra2017efficient}, including challenging flow scenarios as those during search-and-rescue operations at sea \cite{serra2019search}.

\subsection{Three-Dimensional ABC Flow}\label{sec:3dabc}
In this section, iLESs are applied to a fully coupled three-dimensional flow. Additionally, the convergence of the attraction rate and higher-order approximation to the backward-time FTLE field is demonstrated. For this section, the Arnold-Beltrami-Childress (ABC) flow \cite{Arnold2014sur,dombre1986chaotic} was chosen, a divergence-free flow commonly used in FTLE and LCS demonstrations. The fluid components of the ABC flow are analytically given by, 
\begin{equation}
\begin{split}
    \dot x = u &= A\sin(z) + C\cos(y),\\
    \dot y = v &= B\sin(x) + A\cos(z),\\
    \dot z = w &= C\sin(y) + B\cos(x).
    \label{abcflow}
\end{split}
\end{equation}
The ABC flow $\mathbf{v}=(u,v,w)$ is an exact steady solution to Euler's fluid equations and has been shown to have chaotic particle trajectories \cite{dombre1986chaotic}. The domain for $\mathbf{x}=(x,y,z)$ is the periodic cube, $U=[0, 2\pi]\times [0, 2\pi]\times [0, 2\pi]$.
For coefficients,
$A = \sqrt{3},\ B = \sqrt{2},\ C = 1$, were chosen following \cite{henon1966topologie}.

As the ABC flow is an analytical flow, it is possible to analytically express the repulsion and attraction rate fields, as given in Appendix \ref{ABCflow_example_details}.

\begin{figure}[!h]
    \centering
    \includegraphics[width=0.6\linewidth]{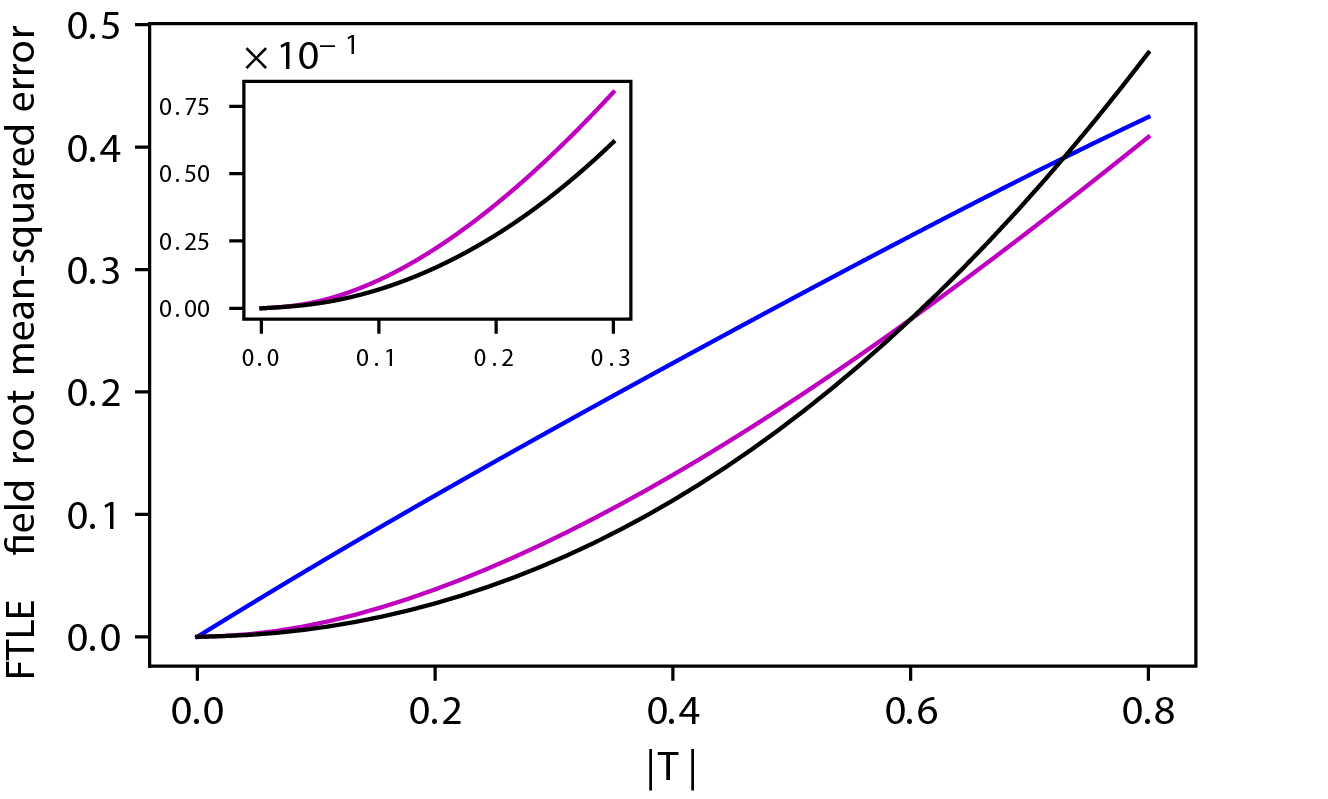}
    \caption{RMSE for successive approximations of the backward-time FTLE field for the ABC flow  \eqref{abcflow} expanded in $T$: zeroth-order (blue), first-order (magenta). Time is in non-dimensional units. The inset shows the behavior for the higher order terms for $|T|$ close to $0$.}
    \label{fig:abc_corr_ts}
\end{figure}

Fig.\ \ref{fig:abc_corr_ts} shows the RMSE of Eulerian approximations with the benchmark FTLE field (computed using the algorithm of \cite{DuMa2010}) as the flow is integrated backward in time from the initial time $t_{0}$.
Since this flow is autonomous, $t_{0}$ is arbitrary. The RMSE for the attraction rate is shown in blue, the first-order approximation in magenta, and the second-order approximation in black. As in the previous sections, this figure shows that as the integration time goes to zero, the RMSE goes to zero as well.

\begin{figure}[h]
    \centering
    \includegraphics[width=\linewidth]{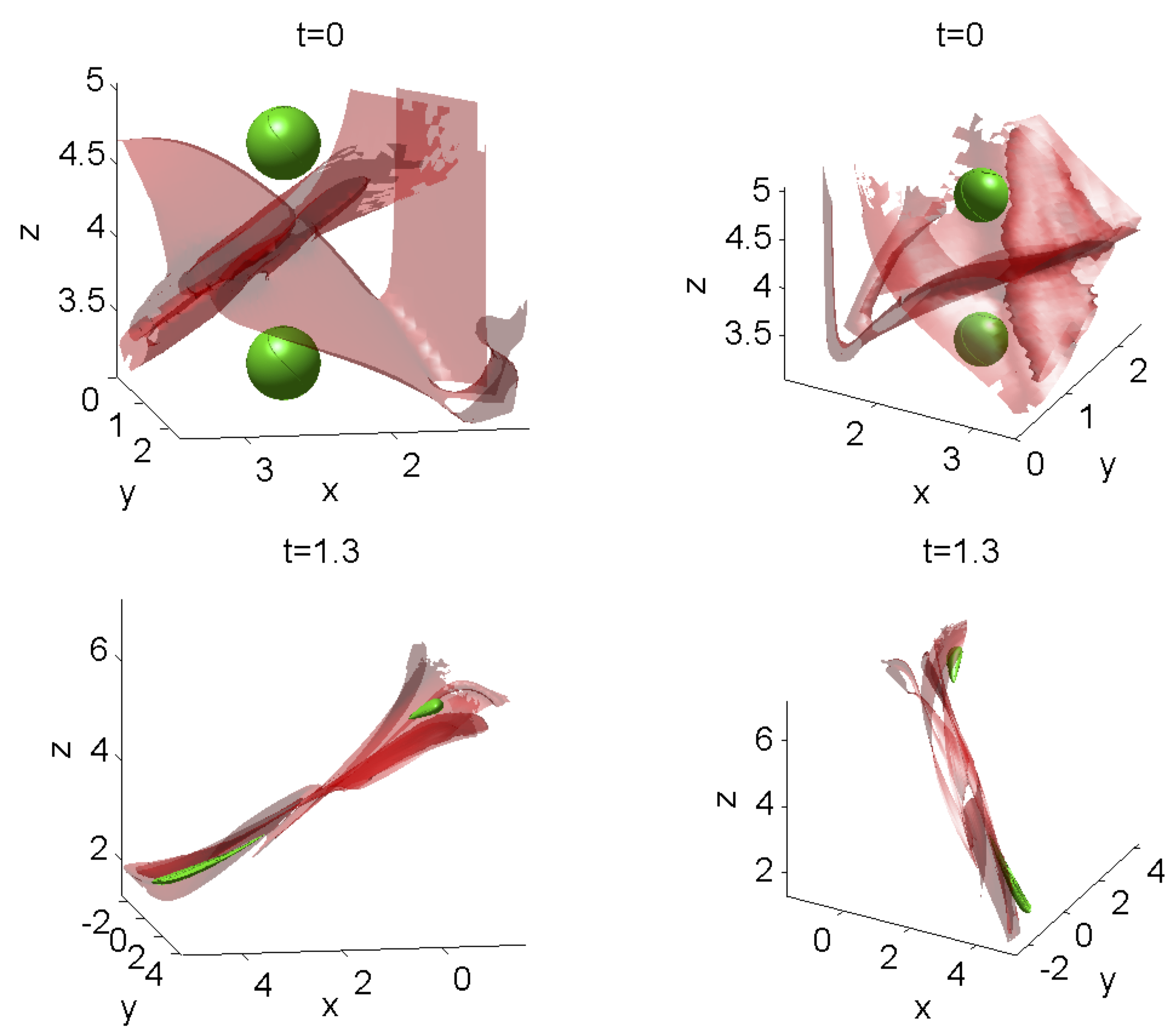}
    \caption{A repelling two-dimensional iLES surface, red, within the three-dimensional flow, with a blob a passive tracers, green, shown at different angles. Top row shows the iLES and tracers at the initial time, $t_0=0$. Bottom row shows the iLES and tracers after being advected forward in time to $t=1.3$.}
    \label{fig:abc_rilcs}
\end{figure}

Fig. 
\ref{fig:abc_rilcs} examines the efficacy of iLESs for the ABC flow. 
Fig.\ \ref{fig:abc_rilcs} shows a repelling iLES (red), along with two blobs of passive tracers (green). 
In this figure, the first row shows the  initial configuration from two different angles, while the second row shows the configuration after being advected by the flow for a time of 1.3 non-dimensional units. 
Due to the the large amounts of twisting and shear in the ABC flow, the repelling effects of iLES are more difficult to visualize in this flow than in the examples of sections \ref{sec:wrf} and \ref{sec:2ddg}. 
To compensate for this, two blobs were used in Fig.\ \ref{fig:abc_rilcs}.
The green blobs are initialized above and below a repelling iLES. 
In this figure one can see that as the iLES and tracers are advected by the flow, the tracer blobs are transported away from each other. This also demonstrates the effectiveness of iLES as an indicator of flow separatrices, as tracers on opposite sides of the iLES do not interact with one another.

\section{Conclusions and Future Directions}\label{sec:conclusion}

Inspired by the recent variational theory of OECSs \cite{serra2016objective} relating Eulerian quantities to short-term Lagrangian transport, this paper provides a connection between the smallest and largest eigenvalues of the Eulerian rate-of-strain tensor and the backward-time and forward-time FTLE fields in $n$ dimensional systems. It was proven that these 
eigenvalues are the limits of the backward-time and forward-time FTLE fields, respectively, as the integration time, $T$, goes to $0$. Additionally, it has shown that for small integration times $|T|$, the eigenvectors of the right Cauchy-Green strain tensor are equal to those of the rate-of-strain tensor. These results provided a new Eulerian diagnostic, iLES, the instantaneous Lyapunov exponent structure, which identifies the major hyperbolic features dominating short-time particle deformation patterns. Therefore, in the same way OECSs are related to LCSs, iLES can be used in the place of FTLE ridges when studying 
flows, and with considerable computational savings. 

We explored our results on several analytical and numerical data sets, demonstrating their efficacy in revealing material transport structure. Given their connection to the widely used FTLE-ridge-based hyperbolic LCS, the iLE and iLES approach could  have wide application.

Moreover, higher-order approximations of the FTLE field using Rivlin-Ericksen tensors were derived and explored. While this study expanded the right Cauchy-Green deformation tensor and FTLE fields to third order and second order in $T$, respectively, one could follow this procedure to arbitrary order $k$, assuming the underlying vector field is smooth enough in the sense of differentiability.  
Automatic differentiation techniques can be utilized, such as used for invariant manifold estimation, where expansions up to order $k=35$ or higher in the dependent variables have been realized \cite{gomez2004connecting,farres2010high,KoLoMaRo2011}.
We note, however, that approximations beyond linear order require more than one time point to be computed from numerical data sets. As a corollary, one could measure the effect of unsteadiness, the $t_0$ dependence, on the FTLE (and corresponding Lagrangian transport structure) by calculating the difference between a high-order FTLE approximation and the true (particle advection based) FTLE.

Future work on this topic will explore: the existence of lower-dimensional iLESs embedded within higher-dimensional iLESs \cite{nave2019global,Doedel_2015,gomez2004connecting,gabern2005theory,surana2008exact,ross2018experimental}; the application of iLESs and higher-order FTLE approximations to experimental data \cite{nolan2018coordinated,nolan2019method}; the application of higher-order FTLE approximations to reduced order models (ROMs) \cite{xie2019lagrangian}; measures of the influence of (temporal) unsteadiness compared with (spatial) inhomogeneity on Lagrangian transport structure; and the determination of the time interval over which Eulerian diagnostics are most effective.

\vspace{10pt} 

\noindent{\small\textbf{Conflict of interest}: The authors declare that they have no conflict of interest.}

\vspace{10pt}

\noindent{\small\textbf{Funding}: This research was supported in part by grants from the National Science Foundation (NSF) under grant numbers AGS 1520825 (Hazards SEES: Advanced Lagrangian Methods for Prediction, Mitigation and Response to Environmental Flow Hazards) and DMS 1821145 (Data-Driven Computation of Lagrangian Transport Structure in Realistic Flows). Any opinions, findings, and conclusions or recommendations expressed in this material are those of the authors and do not necessarily reflect the views of the sponsors.
}

\vspace{10pt}

\noindent{\small\textbf{Acknowledgements}:
We thank Hosein Foroutan for providing us with WRF model data, Nicole Abaid for assistance with the perturbation analysis,  Gary Nave for fruitful conversations related to some of the ideas covered in this work during their nascent stage, and Siavash Ameli for developing and maintaining the TRACE server used for the geophysical particle integration calculations. M.S. acknowledges support from the Schmidt Science Fellowship and the Postdoc Mobility Fellowship from the Swiss National Science Foundation.
}

\vspace{10pt}

\noindent{\bf Appendices}

\vspace{-0.2in}

\begin{appendices}

\section{Expansion of the right Cauchy-Green tensor in the integration time}\label{appendix}


For tensor fields in what follows, the dependence on $\mathbf{x}_0$ and $t_0$ will be notationally dropped for clarity, as it will be understood. For small integration time $T=t-t_0$, the right Cauchy-Green tensor, $\mathbf{C}$,  may be expanded, as in \cite{serra2016objective,nave2018trajectory,truesdell1965handbuch}, in terms of the integration time $T$,
\begin{equation}
\mathbf{C}   = \mathbf{C}|_{T=0} +\left.\frac{d\mathbf{C}}{dT}\right|_{T=0}T + \frac{1}{2!} \left.\frac{d^2\mathbf{C}}{dT^2}\right|_{T=0}T^2+ 
\frac{1}{3!} \left.\frac{d^3\mathbf{C}}{dT^3}\right|_{T=0}T^3+ \mathcal{O}(T^4).
\label{eq:theo:CG Expansion}	
\end{equation}
where the dependence on the initial position and time is understood. Because all derivatives are evaluated at $T=0$, $\left.\tfrac{d}{dt}\right|_{t=t_0}=\left.\tfrac{d}{dT}\right|_{T=0}$. The first term on the right denotes the situation of no deformation, therefore,
$\mathbf{C}|_{T=0}=\mathbbm{1}$. The derivatives of the right Cauchy-Green tensor are given to any order by the Rivlin-Ericksen tensors \cite{nave2018trajectory,truesdell1965handbuch,truesdell2004non},
\begin{equation}
\begin{aligned}
\frac{d\mathbf{C}}{dt} &= \nabla\frac{d\mathbf{x}}{dt}+\left(\nabla\frac{d\mathbf{x}}{dt}\right)^{\hspace{-1mm}\top}, & {~} \\
\frac{d^k\mathbf{C}}{dt^k} &= \nabla\frac{d^k\mathbf{x}}{dt^k} + \left(\nabla\frac{d^k\mathbf{x}}{dt^k}\right)^{\hspace{-1mm}\top}
 + \sum_{i=1}^{k-1}\left(\!\begin{array}{c}
k \\ i
\end{array}\!\right)\left(\nabla\frac{d^i\mathbf{x}}{dt^i}\right)^{\hspace{-1mm}\top} \nabla\frac{d^{k-i}\mathbf{x}}{dt^{k-i}}, &k \geq 2.
\end{aligned}
\end{equation}
For small $|T|\ll1$, the leading order behavior is given by the first Rivlin-Ericksen tensor $(\nabla\mathbf{v}+(\nabla\mathbf{v})^{\top})$, which is twice $\mathbf{S}$, from \eqref{S_matrix}. The second-order term is,
\begin{equation}
\begin{aligned}
\frac{d^2\mathbf{C}}{dt^2} &= \nabla\frac{d^2\mathbf{x}}{dt^2} + \left(\nabla\frac{d^2\mathbf{x}}{dt^2}\right)^{\hspace{-1mm}\top}
 + 
 2 \left(\nabla\frac{d\mathbf{x}}{dt}\right)^{\hspace{-1mm}\top} \nabla\frac{d\mathbf{x}}{dt}, \\
 &= \nabla\frac{d\mathbf{v}}{dt} + \left(\nabla\frac{d\mathbf{v}}{dt}\right)^{\hspace{-1mm}\top}
 +  2 \left(\nabla\mathbf{v}\right)^{\hspace{-1mm}\top} \nabla\mathbf{v},
 \\
  &= 2 \mathbf{B}
\end{aligned}
\end{equation}
where $\mathbf{B}$ is the same as given in \eqref{B_matrix}.

The third-order term is,
\begin{equation}
\begin{aligned}
\frac{d^3\mathbf{C}}{dt^3} 
&= \nabla\frac{d^3\mathbf{x}}{dt^3} + \left(\nabla\frac{d^3\mathbf{x}}{dt^3}\right)^{\hspace{-1mm}\top}
 + 3 \left[ 
 \left(\nabla\frac{d\mathbf{x}}{dt}\right)^{\hspace{-1mm}\top} \nabla\frac{d^{2}\mathbf{x}}{dt^{2}} 
 + 
 \left(\nabla\frac{d^{2}\mathbf{x}}{dt^{2}}\right)^{\hspace{-1mm}\top}
 \nabla\frac{d\mathbf{x}}{dt}
 \right],
 \\
 &= \nabla\frac{d\mathbf{a}}{dt} + \left(\nabla\frac{d\mathbf{a}}{dt}\right)^{\hspace{-1mm}\top}
 + 3 \left[ 
 \left(\nabla\mathbf{v}\right)^{\hspace{-1mm}\top} \nabla\mathbf{a} 
 + 
 \left(\nabla\mathbf{a}\right)^{\hspace{-1mm}\top} \nabla\mathbf{v} 
 \right],
  \\
  &= 3 \mathbf{Q},
\end{aligned}
\end{equation}
where $\mathbf{Q}$ is the same as given in \eqref{Qmatrix}.

The expansion of the right Cauchy-Green tensor (\ref{eq:theo:CG Expansion}) can be written as,
\begin{equation}
\begin{aligned}
\mathbf{C}  
&= \mathbbm{1} + 2T\mathbf{S} 
+ T^2 \mathbf{B} + \tfrac{1}{2}T^3 \mathbf{Q} 
+ \mathcal{O}(T^4),
\\
&=\mathbbm{1} + 2T \left( \mathbf{S} 
+ \tfrac{1}{2}T \mathbf{B} + (\tfrac{1}{2}T)^2 \mathbf{Q}  + \mathcal{O}(T^3) \right),
\end{aligned}
\label{eq:theo:Expansion2}
\end{equation}
which is a form convenient for matrix perturbation analysis, as in Appendix \ref{matrix_perturbation}.

\section{Details of approximating the FTLE to second-order in integration time}\label{expansion_details}

Note the following general result for the eigenvalues, 
\begin{equation}
    \lambda_{-}(\mathbf{A})=\lambda_{1}(\mathbf{A})\leq...\leq\lambda_{n}(\mathbf{A})=\lambda_{+}(\mathbf{A}),
\end{equation}
of $n \times n$ real symmetric matrices $\mathbf{A}$.  
Here, we use $\lambda_{-}(\mathbf{A})$ and $\lambda_{+}(\mathbf{A})$ as shorthand for $\lambda_{\rm min}(\mathbf{A})$ and $\lambda_{\rm max}(\mathbf{A})$, the minimum and maximum eigenvalues of $\mathbf{A}$, respectively.
For scalar $c \ne 0$,
\begin{equation}
\lambda_{\pm}(\mathbbm{1} + c \mathbf{A})
=
1 + \lambda_{\pm}(c \mathbf{A}),
\label{eigenvalue_equality}
\end{equation}
where,
\begin{equation}
   \lambda_{\pm}(c \mathbf{A}) =
    \begin{cases}
      c \lambda_{\pm}(\mathbf{A}), & \text{for}\ c>0, \\
      c \lambda_{\mp}(\mathbf{A}), & \text{for}\ c<0.
    \end{cases}
\end{equation}
See Appendix \ref{proof_eq} for the proof.

In \eqref{FTLE},  $\lambda_{n} = \lambda_{+}(\mathbf{C}_{t_{0}}^{t}(\mathbf{x}))$. For small $T>0$, where the $\mathcal{O}(T^2)$ and higher terms can be neglected,
\begin{equation}
\lambda_{+}(\mathbf{C}_{t_{0}}^{t}(\mathbf{x})) = 
1 + 2T\lambda_{+}(\mathbf{S}(\textbf{x},t_0)) + \mathcal{O}(T^2).
\label{Cauchy_Green_max_eig}
\end{equation}
Thus,
\begin{equation}
\log(\lambda_n) = \log(1 + 2T\lambda_{+}(\mathbf{S}(\textbf{x},t_0)))=
2T\lambda_{+}(\mathbf{S}(\textbf{x},t_0))
= 2Ts_n(\textbf{x},t_0),
\end{equation}
in the limit of small $T$ using the Taylor expansion, $\log(1+\delta) = \delta + \mathcal{O}(\delta^2)$ for small $|\delta|$.

From \eqref{FTLE}, and noting that $|T|=T$ for $T>0$,
\begin{equation}
    \sigma_{t_{0}}^{t}(\mathbf{x}) = \frac{1}{2|T|}\log(\lambda_{n})
    =\frac{1}{2T}2Ts_n (\textbf{x},t_0)
    =s_n(\textbf{x},t_0)
\end{equation}
Therefore, the maximum eigenvalue of $\mathbf{S}(\textbf{x},t_0)$ is the limit of the FTLE value for forward time as $T \rightarrow 0^{+}$.

For $T<0$ with small $T$,
\begin{equation}
\lambda_{+}(\mathbf{C}_{t_{0}}^{t}(\mathbf{x})) = 
1 + 2T\lambda_{-}(\mathbf{S}(\textbf{x},t_0))  + \mathcal{O}(T^2).
\label{eq:theo:eig_value_expansion}
\end{equation}
Thus,
\begin{equation}
\log(\lambda_n) = 2T\lambda_{-}(\mathbf{S}(\textbf{x},t_0))
= 2Ts_1(\textbf{x},t_0),
\end{equation}
in the limit of small $T$.

From \eqref{FTLE}, and noting that $|T|=-T$ for $T<0$,
\begin{equation}
    \sigma_{t_{0}}^{t}(\mathbf{x}) = \frac{1}{2|T|}\log(\lambda_{n})
    =-\frac{1}{2T}2Ts_1(\textbf{x},t_0)
    =-s_1(\textbf{x},t_0).
\end{equation}
Therefore, the negative of the minimum eigenvalue of $\mathbf{S}(\textbf{x},t_0)$ is the limit of the FTLE value for backward time as $T \rightarrow 0^{-}$.

Consider now the third term, the order $T^2$ term, in the expansion \eqref{Cauchy_Green_expansion} of the right Cauchy-Green tensor. Then \eqref{Cauchy_Green_max_eig} becomes,
\begin{equation}
\lambda_{+}(\mathbf{C}_{t_{0}}^{t}(\mathbf{x})) = 
1 + 2T\lambda^{+}\Big(\mathbf{S}(\textbf{x},t_0) 
+\tfrac{1}{2}T \mathbf{B}(\textbf{x},t_0) 
\Big) + \mathcal{O}(T^3).
\label{Cauchy_Green_max_eig_correction}
\end{equation}
Note that $\mathbf{B}(\textbf{x},t_0)$, like $\mathbf{S}(\textbf{x},t_0)$, is symmetric. 

Below, we adopt the notation of $s_-$ and $s_+$ for $s_1$ and $s_n$, respectively, as in the main text.

It can be shown via matrix perturbation techniques (see Appendix \ref{matrix_perturbation}) that, 
\begin{equation}
\lambda_{+}\Big(\mathbf{S}(\textbf{x},t_0) 
+\tfrac{1}{2}T \mathbf{B}(\textbf{x},t_0) 
\Big) = s_+ + \tfrac{1}{2}T \bm{e}_{+}^{\top} \mathbf{B} \bm{e}_{+} + \mathcal{O}(T^2).
\label{eigenvalue_perturbation}
\end{equation}
Using the Taylor expansion $\log(1+\delta) = \delta -\tfrac{1}{2}\delta^2 + \tfrac{1}{3}\delta^3 +\mathcal{O}(\delta^4)$ for small $|\delta|$,
by a similar argument as before, for small $T$,
\begin{equation}
\begin{split}
\log(\lambda_{+}(\mathbf{C}_{t_{0}}^{t}(\mathbf{x}))) 
&= \log \Big(1 + 2T \Big[ s_+ + \tfrac{1}{2}T \bm{e}_{+}^{\top} \mathbf{B} \bm{e}_{+}  + \mathcal{O}(T^2) \Big] \Big), \\
&= 2T \Big[ s_+ + \tfrac{1}{2}T \bm{e}_{+}^{\top} \mathbf{B} \bm{e}_{+} + \mathcal{O}(T^2) \Big]
- \tfrac{1}{2} 4T^2 s_+^2 + \mathcal{O}(T^3), \\
&= 2T \Big[ s_+ + T \Big( - s_+^2 + \tfrac{1}{2} \bm{e}_{+}^{\top} \mathbf{B} \bm{e}_{+} \Big) + \mathcal{O}(T^2) \Big].
\end{split}
\end{equation}
Therefore, for $T>0$ with small $|T|$,
\begin{equation}
    \sigma_{t_{0}}^{t}(\mathbf{x}) = s_+(\textbf{x},t_0)
    + T \Big( - s_+(\textbf{x},t_0)^2 + \tfrac{1}{2} \bm{e}_{+}(\textbf{x},t_0)^{\top} \mathbf{B}(\textbf{x},t_0) \bm{e}_{+}(\textbf{x},t_0) \Big)
    + \mathcal{O}(T^2).
    \label{FTLE_T_positive_correction}
\end{equation}
And similarly, for  $T<0$ with small $|T|$,
\begin{equation}
    \sigma_{t_{0}}^{t}(\mathbf{x}) = -s_-(\textbf{x},t_0)
    - T \Big(- s_-(\textbf{x},t_0)^2 + \tfrac{1}{2} \bm{e}_{-}(\textbf{x},t_0)^{\top} \mathbf{B}(\textbf{x},t_0) \bm{e}_{-}(\textbf{x},t_0) \Big)
    + \mathcal{O}(T^2).
    \label{FTLE_T_negative_correction}
\end{equation}
If we continue this procedure to obtain the approximate  forward and backward FTLE through second order in $T$, we get \eqref{FTLE_T_correction2}.

\section{Proof of equation \eqref{eigenvalue_equality}}\label{proof_eq}

Let $\mathbf{A}$ be an $n \times n$ matrix, $\lambda$ an eigenvalue of $\mathbf{A}$, $\bm{\xi}$ the corresponding eigenvector of $\mathbf{A}$, $\mathbbm{1}$  the $n \times n$ identity matrix and $\zeta \in \mathbb{C}$. By the definition of an eigenvalue $\mathbf{A}\bm{\xi}$ = $\lambda\bm{\xi}$, we have,
\begin{equation}
\left(\zeta\,\mathbbm{1}+\mathbf{A}\right)\bm{\xi} = \zeta\,\bm{\xi}+\mathbf{A}\bm{\xi}=\zeta\,\bm{\xi}+\lambda\bm{\xi}=\left(\zeta+\lambda\right)\bm{\xi}.    
\end{equation}
Therefore, if $\lambda$ is an eigenvalue of $\mathbf{A}$ with eigenvector $\bm{\xi}$, then $(\zeta+\lambda)$ is an eigenvalue of $\zeta\,\mathbbm{1}+\mathbf{A}$ with the same eigenvector $\bm{\xi}$.  In particular, this holds when $\zeta=1$, as in \eqref{eigenvalue_equality}.

\section{Equality of the eigenvectors of $S$ and $C$ as integration time goes to zero}\label{equality_eigenvectors}

Let $T>0$ be small enough that the relationships in \eqref{Cauchy_Green_expansion} and \eqref{Cauchy_Green_max_eig} hold and $\mathcal{O}(T^2)$ terms are negligible.
As before, let $\bm{e}_{i}$ be the eigenvector of $\mathbf{S}$ associated with $s_{i}$, then,
\begin{align}
    \mathbf{S}\,\bm{e}_{i}&=s_{i}\bm{e}_{i}\label{eq:theo:s_vec},\\
    2 T \mathbf{S}\,\bm{e}_{i}+\bm{e}_{i}&= 2 T s_{i}\bm{e}_{i}+\bm{e}_{i},\\
    \left(2 T \mathbf{S}+\mathbbm{1}\right)\bm{e}_{i}&= \left(2 T s_{i}+1\right)\bm{e}_{i},\\
    \mathbf{C}\,\bm{e}_{i}&=\lambda_{i}\bm{e}_{i}\label{eq:theo:c_vec},
\end{align}
where the dependence on $\mathbf{x}$ and $t_0$ is understood and we used the order-$T$ approximation for $\mathbf{C}$.
But from \eqref{C_eigvals} and \eqref{C_eigvecs}, 
\begin{align}
\mathbf{C}\,\bm{\xi}_{\lambda_i}=\lambda_{i}\bm{\xi}_{\lambda_i}
\end{align}
thus,
\begin{align}
\bm{e}_i = \bm{\xi}_{\lambda_i}
\end{align}
that is, if $\bm{e}_{i}$ is an eigenvector of $\mathbf{S}$, then 
it is also an eigenvector of $\mathbf{C}$ in the limit as $T\rightarrow 0$. Now, assuming that $\bm{\xi}_{\lambda_i}$ is the eigenvector of $\mathbf{C}$ associated with $\lambda_{i}$, and working through  (\ref{eq:theo:s_vec}-\ref{eq:theo:c_vec}) in reverse proves that if $\bm{\xi}_{\lambda_i}$ is an eigenvector of $\mathbf{C}$, then 
it is also an eigenvector of $\mathbf{S}$ in the limit as $T$ goes to $0$. For $T<0$ an analogous argument holds using \eqref{eq:theo:eig_value_expansion} in place of \eqref{Cauchy_Green_max_eig} and with the ordering of the eigenvalues opposed, i.e. $\lambda_i \sim s_{n-i+1},\,i\in\{1,\ldots,n\}$.


Therefore, in the limit as $|T|$ goes to $0$, the eigenvectors of $\mathbf{S}$ and $\mathbf{C}$ are equal.  For small $|T|$, we can also use the perturbation expansion of $\mathbf{C}$, to get the estimated eigenvectors of $\mathbf{C}$ from \eqref{eigvec_expand} in Appendix \ref{matrix_perturbation}, which provides  the eigenvectors through order $T^2$ using only the velocity field $\mathbf{v}$ from \eqref{eq:theo:1} evaluated at $\mathbf{x}$ and time $t_0$ as well as appropriate derivatives.

\section{Eigenvalues of the Taylor-expanded right Cauchy-Green tensor}\label{matrix_perturbation}

Let $\mathbf{S}$ be a real, symmetric $n \times n$ matrix with $n$ distinct eigenvalues, and let $\mathbf{B}$ and $\mathbf{Q}$ also be real, symmetric $n \times n$ matrices.  We seek the eigenvalues of, 
\begin{equation}
\mathbf{S}_{\varepsilon} = \mathbf{S} + \varepsilon\mathbf{B} + \varepsilon^2\mathbf{Q},
\end{equation}
a perturbation of $\mathbf{S}$, where $|\varepsilon|$ is a small scalar. In our case of interest, from \eqref{eq:theo:Expansion2}, the small parameter is $\varepsilon = \tfrac{1}{2}T$.

Consider the eigenvalue $\mu_0$ of $\mathbf{S}$ with corresponding normalized eigenvector $\bm{\xi}_0$.  Let's refer to the perturbed eigenvalue and corresponding perturbed eigenvector of $\mathbf{S}_{\varepsilon}$ as $\mu_{\varepsilon}$ and $\bm{\xi}_{\varepsilon}$.  One can expand 
$\bm{\xi}_{\varepsilon}$ and $\mu_{\varepsilon}$ in powers of $\varepsilon$ as
\begin{align}
\bm{\xi}_{\varepsilon} &= \bm{\xi}_{0} + \varepsilon \bm{\xi}_{1} +
\varepsilon^2 \bm{\xi}_{2} + \mathcal{O}(\varepsilon^3),  \label{eigvec_expand}\\
\mu_{\varepsilon} &= \mu_{0} + \varepsilon \mu_{1} +
\varepsilon^2 \mu_{2} + \mathcal{O}(\varepsilon^3). \label{eigval_expand}
\end{align}
The eigenvector equation, $\mathbf{S}_{\varepsilon}\bm{\xi}_{\varepsilon} = \mu_{\varepsilon} \bm{\xi}_{\varepsilon}$,
can be approximated as
\begin{equation}
(\mathbf{S} + \varepsilon\mathbf{B} + \varepsilon^2\mathbf{Q}) (\bm{\xi}_{0} + \varepsilon \bm{\xi}_{1} + \varepsilon^2 \bm{\xi}_{2}) = (\mu_{0} + \varepsilon \mu_{1} + \varepsilon^2 \mu_{2})(\bm{\xi}_{0} + \varepsilon \bm{\xi}_{1} + \varepsilon^2 \bm{\xi}_{2}),
\end{equation}
which leads to the following three expressions, corresponding to the order one terms, order $\varepsilon$, and order $\varepsilon^2$ terms, respectively,
\begin{align}
\mathbf{S} \bm{\xi}_{0} &= \mu_{0} \bm{\xi}_{0}, \label{order_one}\\
\mathbf{S} \bm{\xi}_{1} + \mathbf{B} \bm{\xi}_{0} &= 
\mu_{0} \bm{\xi}_{1} + \mu_1 \bm{\xi}_{0}, \label{order_epsilon}\\
\mathbf{S} \bm{\xi}_{2}+\mathbf{B} \bm{\xi}_{1}+\mathbf{Q}\bm{\xi}_{0} 
&= 
\mu_{0} \bm{\xi}_{2} + \mu_1 \bm{\xi}_{1} + \mu_2 \bm{\xi}_{0}. \label{order_epsilonsquared}
\end{align}
Multiply \eqref{order_epsilon} by $\bm{\xi}_{0}^{\top}$ to get,
\begin{equation}
\bm{\xi}_{0}^{\top}\mathbf{S} \bm{\xi}_{1} + \bm{\xi}_{0}^{\top}\mathbf{B} \bm{\xi}_{0} = 
\mu_{0} \bm{\xi}_{0}^{\top}\bm{\xi}_{1} + \mu_1 \bm{\xi}_{0}^{\top}\bm{\xi}_{0}, \label{eig_pert}
\end{equation}
Since $\bm{\xi}_{0}$ is normalized, $\bm{\xi}_{0}^{\top}\bm{\xi}_{0}=1$.  Also, since $\mathbf{S}$ is symmetric,
\begin{equation}
    \begin{split}
        \bm{\xi}_{0}^{\top}\mathbf{S} \bm{\xi}_{1} &= 
        (\bm{\xi}_{1}^{\top}\mathbf{S} \bm{\xi}_{0})^{\top} , \\
        &= (\bm{\xi}_{1}^{\top}\mu_0 \bm{\xi}_{0})^{\top} , \\
        &= \mu_0 \bm{\xi}_{0}^{\top}\bm{\xi}_{1},
    \end{split}
\end{equation}
where \eqref{order_one} was used. Now \eqref{eig_pert} is,
\begin{equation}
\mu_0 \bm{\xi}_{0}^{\top}\bm{\xi}_{1} + \bm{\xi}_{0}^{\top}\mathbf{B} \bm{\xi}_{0} = 
\mu_{0} \bm{\xi}_{0}^{\top}\bm{\xi}_{1} + \mu_1. \label{eig_pert2}
\end{equation}
Thus,
\begin{equation}
\mu_1 = \bm{\xi}_{0}^{\top}\mathbf{B} \bm{\xi}_{0},
\label{eig_pert3}
\end{equation}
which, since $\mathbf{B}$ is symmetric, represents a quadratic form.

A bound can be put on the term $\bm{\xi}_{0}^{\top}\mathbf{B} \bm{\xi}_{0}$, noting that $\bm{\xi}_{0}$ is a unit vector.  If $b_n$ is the maximum eigenvalue of $\mathbf{B}$, then,
\begin{equation}
    \max_{\bm{\xi}_{0}} \bm{\xi}_{0}^{\top}\mathbf{B} \bm{\xi}_{0} = b_n.
\end{equation}
Similarly, if $b_1$ is the minimum eigenvalue of $\mathbf{B}$, then,
\begin{equation}
    \min_{\bm{\xi}_{0}} \bm{\xi}_{0}^{\top}\mathbf{B} \bm{\xi}_{0} = b_1.
\end{equation}
So,
\begin{equation}
    \mu_1 = \bm{\xi}_{0}^{\top}\mathbf{B} \bm{\xi}_{0} \in [b_1,b_n].
\end{equation}
So \eqref{eigval_expand} becomes,
\begin{equation}
\mu_{\varepsilon} = \mu_0 + \varepsilon \mu_1 + \mathcal{O}(\varepsilon^2),
\label{eig_pert4}
\end{equation}
where $\mu_1$ is from \eqref{eig_pert3}.

With $\mu_1$ in hand, $\bm{\xi}_{1}$ can also be determined as the solution of the following re-arranged version of \eqref{order_epsilon},
\begin{equation}
(\mathbf{S} - \mu_0 \mathbbm{1})\bm{\xi}_{1} = - (\mathbf{B} - \mu_1 \mathbbm{1}) \bm{\xi}_{0}.
\label{xi1}
\end{equation}
Note that $(\mathbf{S} - \mu_0 \mathbbm{1})$ is not invertible as it has zero determinant, since $\mu_0$ is an eigenvalue of $\mathbf{S}$. The null space of $(\mathbf{S} - \mu_0 \mathbbm{1})$ is span$\{\bm{\xi}_{0}\}$. Note that \eqref{xi1} is of the form $\mathbf{A}\mathbf{x}=\mathbf{b}$ with a square matrix $\mathbf{A}$ of nullity 1 and a vector $\mathbf{b}$ which is in the image of $\mathbf{A}$, as shown below.

Note that, as a consequence of \eqref{eig_pert3}, the vector  $\mathbf{B}\bm{\xi}_{0}$ can be written as,
\begin{equation}
     \mathbf{B}\bm{\xi}_{0}= \mu_1 \bm{\xi}_{0} + d  \bm{\xi}_{0}^{\prime \perp},
     \label{Bxi0}
\end{equation}
where 
$d \in \mathbbm{R}$ and $\bm{\xi}_{0}^{\prime\perp}$ is, in general, 
a vector in ${\rm im}(\mathbf{S} - \mu_0 \mathbbm{1})$.
This equation can be re-arranged to yield,
\begin{equation}
 - \mathbf{B}\bm{\xi}_{0} + \mu_1 \bm{\xi}_{0} 
 =-d  \bm{\xi}_{0}^{\prime \perp}.
\label{xi1a}
\end{equation}
The left-hand side of \eqref{xi1a} is the same as the right-hand side of \eqref{xi1},
which means the right-hand side of \eqref{xi1} is a vector $\mathbf{b}$ which is in ${\rm im}(\mathbf{S} - \mu_0 \mathbbm{1})$, which we will use below. 

One can 
determine $\mu_2$ by multiplying \eqref{order_epsilonsquared} by $\bm{\xi}_{0}^{\top}$ to get, by a similar procedure as before,
\begin{equation}
\mu_0 \bm{\xi}_{0}^{\top} \bm{\xi}_{2}+\bm{\xi}_{0}^{\top}\mathbf{B} \bm{\xi}_{1}+\bm{\xi}_{0}^{\top}\mathbf{Q}\bm{\xi}_{0} 
= 
\mu_{0} \bm{\xi}_{0}^{\top}\bm{\xi}_{2} + \mu_1 \bm{\xi}_{0}^{\top}\bm{\xi}_{1} + \mu_2.
\end{equation}
Canceling the identical terms on both sides, we get,
\begin{equation}
\mu_2=
\bm{\xi}_{0}^{\top}\mathbf{Q}\bm{\xi}_{0} +\bm{\xi}_{0}^{\top}\mathbf{B} \bm{\xi}_{1}
- \mu_1 \bm{\xi}_{0}^{\top}\bm{\xi}_{1}. 
\label{lambda_2}
\end{equation}
But take the transpose and, 
\begin{eqnarray}
\mu_2 &=&
\bm{\xi}_{0}^{\top}\mathbf{Q}\bm{\xi}_{0} +\bm{\xi}_{1}^{\top}( \mathbf{B} 
- \mu_1 \mathbbm{1}) \bm{\xi}_{0}, 
\\
&=&
\bm{\xi}_{0}^{\top}\mathbf{Q}\bm{\xi}_{0} -\bm{\xi}_{1}^{\top}(\mathbf{S} - \mu_0 \mathbbm{1})\bm{\xi}_{1}, 
\end{eqnarray}
where \eqref{xi1} was used. One can write $\bm{\xi}_{1}$ as,
\begin{equation}
\bm{\xi}_{1} = a\bm{\xi}_{0} + b \bm{\xi}_{0}^{\perp},
\label{xi1_components}
\end{equation}
where $a,b \in \mathbbm{R}$ and $\bm{\xi}_{0}^{\perp} \in {\rm im}(\mathbf{S} - \mu_0 \mathbbm{1})$, which is, in general, not equal to $\bm{\xi}_{0}^{\prime\perp}$ from \eqref{xi1a}. Hence,
\begin{equation}
  \mu_2 = \bm{\xi}_{0}^{\top}\mathbf{Q}\bm{\xi}_{0} 
  -b^2 \bm{\xi}_{0}^{\perp {\top}} (\mathbf{S}  - \mu_0 \mathbbm{1}) \bm{\xi}_{0}^{\perp}.
  \label{lambda2_a}
\end{equation}
Therefore, the only part of $\bm{\xi}_{1}$ which contributes to $\mu_2$ is the part which is in 
${\rm im}(\mathbf{S} - \mu_0 \mathbbm{1})$.

When dealing with a two-dimensional flow field, ${\rm im}(\mathbf{S} - \mu_0 \mathbbm{1})$ is just a 1-dimensional subspace of $\mathbbm{R}^2$, and thus $\bm{\xi}_{0}^{\prime \perp}$ in \eqref{xi1a} is parallel to $\bm{\xi}_{0}^{\perp}$ in \eqref{xi1_components}. Without loss of generality, they can be taken to be equal unit vectors, $\bm{\xi}_{0}^{\perp}=\bm{\xi}_{0}^{\prime \perp}$. Thus, \eqref{xi1} becomes,
\begin{equation}
b(\mathbf{S} - \mu_0 \mathbbm{1})\bm{\xi}_{0}^{\perp} = - d \bm{\xi}_{0}^{\perp},
\label{xi1_eigen_a}
\end{equation}
or, assuming $b\ne0$,
\begin{equation}
(\mathbf{S} - \mu_0 \mathbbm{1})\bm{\xi}_{0}^{\perp} = - \tfrac{d}{b} \bm{\xi}_{0}^{\perp},
\label{xi1_eigen}
\end{equation}
which is an eigenvector equation for the matrix $(\mathbf{S} - \mu_0 \mathbbm{1})$ with the eigenvector $\bm{\xi}_{0}^{\perp}$  and  corresponding eigenvalue $\bar \mu=- \tfrac{d}{b}$. 
Note that if $b=0$, then $d=0$ also, from \eqref{xi1_eigen_a}. 

For two-dimensional flows, from  $\bm{\xi}_{0}$, one can easily obtain $\bm{\xi}_{0}^{\perp}$  from a $90^{\circ}$ counterclockwise rotation,
\begin{equation}
\bm{\xi}_{0}^{\perp} = 
\mathbf{R}\bm{\xi}_{0}, \label{xiperp}
\end{equation}
where,
\begin{equation}
\mathbf{R} =
\begin{bmatrix}
0 & -1 \\
1 & ~~ 0
\end{bmatrix}.
\end{equation}
Now, $\bm{\xi}_{0}^{\perp}$ can be used to obtain $\bar \mu$   from \eqref{xi1_eigen} for the case $d\ne0$. With \eqref{xiperp} in \eqref{xi1_eigen}, \eqref{xi1_eigen} becomes the following eigenvector equation for $\mathbf{R}^{\top}(\mathbf{S} - \mu_0 \mathbbm{1})\mathbf{R}$ with eigenvector $\bm{\xi}_{0}$, 
\begin{equation}
\mathbf{R}^{\top}(\mathbf{S} - \mu_0 \mathbbm{1})\mathbf{R}\bm{\xi}_{0}= \bar  \mu \bm{\xi}_{0},
\label{xi1_eigen_R}
\end{equation}
Therefore $\bar \mu$ is obtained by taking the dot product with $\bm{\xi}_{0}$,
\begin{equation}
\bar \mu = \bm{\xi}_{0}^{\top}\mathbf{R}^{\top}(\mathbf{S} - \mu_0 \mathbbm{1})\mathbf{R}\bm{\xi}_{0},
\label{xi1_eigen_mu}
\end{equation}
and $d$ is obtained from \eqref{xi1a}, noting that $\bm{\xi}_{0}^{\perp {\top}}\bm{\xi}_{0}=0$,
\begin{equation}
\begin{split}
d 
&= \bm{\xi}_{0}^{\perp {\top}} \mathbf{B}\bm{\xi}_{0},\\
&= \bm{\xi}_{0}^{\top}\mathbf{R}^{\top} \mathbf{B}\bm{\xi}_{0}.
\label{xi1_d}
\end{split}
\end{equation}
Thus, \eqref{lambda2_a}, for two-dimensional systems, simplifies to,
\begin{equation}
  \mu_2 = 
  \left\{\begin{array}{@{}ll@{}}
    \bm{\xi}_{0}^{\top}\mathbf{Q}\bm{\xi}_{0}, & \text{if}\ d=0 \\
    \bm{\xi}_{0}^{\top}\mathbf{Q}\bm{\xi}_{0} 
  -\tfrac{d^2}{\bar \mu}, & \text{if}\ d\ne0
  \end{array}\right.
    \label{lambda2_2D}
\end{equation}
where $\bar \mu$ and $d$  are  from  \eqref{xi1_eigen_mu} and \eqref{xi1_d}, respectively.

\section{Details for the examples}\label{example_details}

\subsection{Details for nonlinear saddle example}\label{nonlinear_saddle_details}

Writing the $\log$ term of \eqref{nonlinear_saddle_FTLE} as follows, using Taylor series approximations for small $|T|$, we have,
\begin{equation}
\begin{aligned}
\log & (e^{4T}) - \log[ (1 - y_0^2)e^{2T} + y_0^2)^3], \\
&=
4T - 3 \log[ (1 - y_0^2)(1 + 2T + \tfrac{1}{2!}(2 T)^2 + \tfrac{1}{3!}(2 T)^3 + \mathcal{O}(T^4)) + y_0^2], \\
&=
4T - 3 \log[ 1 + (1 - y_0^2)2T + (1 - y_0^2)2 T^2 + (1 - y_0^2)\tfrac{4}{3}T^3 + \mathcal{O}(T^4) ], \\
&=
4T - 3 [  (1 - y_0^2)2T +y_0^2(1-y_0^2)2T^2 -  y_0^2(1 - y_0^2)(1 - 2y_0^2)\tfrac{4}{3} T^3 + \mathcal{O}(T^4) ], \\
&=
4T - (1 - y_0^2)6T - y_0^2(1 - y_0^2)6T^2 +4T^3y_0^2(1 - y_0^2)(1 - 2y_0^2) + \mathcal{O}(T^4), \\
&=
-2T [(1 - 3y_0^2) + 3 y_0^2 (1 - y_0^2)T - 2y_0^2(1 - y_0^2)(1 - 2y_0^2)T^2 + \mathcal{O}(T^3)].
\end{aligned}
\end{equation}
So the backward FTLE is expanded in $T$ as follows, obtained by dividing by $-2T$, 
\begin{equation}
    \sigma_0^T(\mathbf{x}_0) = (1 - 3y_0^2) + 3  y_0^2 (1 - y_0^2) T - 2y_0^2(1 - y_0^2)(1 - 2y_0^2)T^2 + \mathcal{O}(T^3).
    \label{nonlinear_saddle_FTLE_approximation}
\end{equation}
which is the same as \eqref{nonlinear_saddle_FTLE_approximation_main}.

The FTLE can be approximated by the first, second, and third terms (the zeroth-order, first-order, and second-order in $T$, respectively) using the procedure outlined in section \ref{eigS_T}.
The gradient of the velocity is,
\begin{equation}
\nabla \mathbf{v}(\mathbf{x}_0) =
\begin{bmatrix}
     1 & 0 \\ 
     0 & (- 1 + 3y_0^2)   
\end{bmatrix},
     \label{nonlinear_saddle_gradient}
\end{equation}
which is also $\mathbf{S}(\mathbf{x}_0)$, since the gradient is diagonal. This has  a minimum eigenvalue $s_{-} = - 1 + 3y_0^2$, the negative of which matches the first  term of \eqref{nonlinear_saddle_FTLE_approximation}, as prescribed by \eqref{FTLE_T_correction2}.  To calculate the second term of \eqref{nonlinear_saddle_FTLE_approximation}, the  term first-order in $T$, the acceleration field needs to be calculated and then $\mathbf{B}(\mathbf{x}_0)$.  The acceleration field is, following \eqref{acceleration_field},
\begin{equation}
\begin{split}
\ddot x = \tfrac{d}{dt}\dot x &= x, \\
\ddot y = \tfrac{d}{dt}\dot y &= y - 4 y^3 + 3y^5.
\label{nonlinear_saddle_accel}
\end{split}
\end{equation}
Therefore \eqref{B_matrix} gives,
\begin{equation}
\begin{split}
\mathbf{B}(\mathbf{x}_0) &=
\begin{bmatrix}
     1 & 0 \\ 
     0 & (1 - 12y_0^2 + 15y_0^4)  
\end{bmatrix}
+
\begin{bmatrix}
     1 & 0 \\ 
     0 & (1 - 6y_0^2 + 9y_0^4)  
\end{bmatrix}, \\
&=
\begin{bmatrix}
     2 & 0 \\ 
     0 & (2 - 18y_0^2 + 24y_0^4)  
\end{bmatrix}.
\end{split}
     \label{nonlinear_saddle_B}
\end{equation}
The normalized eigenvector of $\mathbf{S}(\mathbf{x}_0)$ corresponding to $s_-$ is simply $\bm{e}_{-}=[ 0, 1]^{\top}$, which, via \eqref{mu_variables}, yields,
\begin{equation}
\begin{split}
\mu_{1-} = \bm{e}_{-}^{\top} \mathbf{B}(\textbf{x}_0) \bm{e}_{-} 
&=
\begin{bmatrix}
     0 & 1 \\ 
\end{bmatrix}
\begin{bmatrix}
     2 & 0 \\ 
     0 & (2 - 18y_0^2 + 24y_0^4)  
\end{bmatrix}
\begin{bmatrix}
     0 \\ 1  
\end{bmatrix},
\\
&= 2 - 18y_0^2 + 24y_0^4,
\end{split}
     \label{nonlinear_saddle_Bmultiply}
\end{equation}
hence,
\begin{equation}
\begin{split}
-s_-^2 + \tfrac{1}{2} \mu_{1-}  
&=-(1 - 6y_0^2 + 9y_0^4) + 1 -9y_0^2 + 12y_0^4,
\\
&=-3y_0^2 (1 - y_0^2),
\end{split}
     \label{nonlinear_saddle_T_term}
\end{equation}
the negative of which matches the $T$ coefficient of the second term of \eqref{nonlinear_saddle_FTLE_approximation}, as prescribed by \eqref{FTLE_T_correction2}. 

For the term second-order in $T$, note that, as prescribed by \eqref{Qmatrix},
\begin{equation}
\begin{split}
\mathbf{Q}(\mathbf{x}_0) &=
\frac{2}{3}
\begin{bmatrix}
     1 & 0 \\ 
     0 & (-1 + 39y_0^2 - 135 y_0^4 + 105y_0^6)  
\end{bmatrix}\\
&
+
2\begin{bmatrix}
     1 & 0 \\ 
     0 & (1 - 12y_0^2 + 15y_0^4)  
\end{bmatrix}
\begin{bmatrix}
     1 & 0 \\ 
     0 & (- 1 + 3y_0^2)   
\end{bmatrix}, \\
&=
\begin{bmatrix}
     \tfrac{2}{3} & 0 \\ 
     0 & (-\tfrac{2}{3} + 26y_0^2 - 90 y_0^4 + 70y_0^6)  
\end{bmatrix}\\
&+
\begin{bmatrix}
     2 & 0 \\ 
     0 & (-2            + 30y_0^2 - 102y_0^4 + 90y_0^6)
\end{bmatrix}, \\
&=
\begin{bmatrix}
     \tfrac{8}{3} & 0 \\ 
     0 &  (-\tfrac{8}{3} + 56y_0^2 - 192 y_0^4 + 160 y_0^6)   
\end{bmatrix},
     \label{nonlinear_saddle_Q}
\end{split}
\end{equation}
and since \eqref{xi1} implies that $\bm{\xi}_{1-}$ is parallel to $\bm{e}_-$, \eqref{mu_variables} yields,
\begin{equation}
\mu_{2-} = \bm{e}_{-}^{\top} \mathbf{Q}(\textbf{x}_0) \bm{e}_{-}
= -\tfrac{8}{3} + 56y_0^2 - 192 y_0^4 + 160 y_0^6.
\end{equation}
According to \eqref{FTLE_T_correction2}, the second-order term is,
\begin{equation}
\begin{split}
-T^2 &[
-\tfrac{4}{3}(1-3y_0^2)(1 - 6y_0^2 + 9y_0^4) \\
 & ~~~+ (1-3y_0^2) (2 - 18y_0^2 + 24y_0^4)  \\
 & ~~~+ \tfrac{1}{4} (-\tfrac{8}{3} + 56y_0^2 - 192 y_0^4 + 160 y_0^6)]  \\
 = -T^2 &[ 
 ( -\tfrac{4}{3} + 8y_0^2 - 12y_0^4 + 4y_0^2 - 24y_0^4 +36 y_0^6)\\
 & ~~~+ (2 - 18y_0^2 + 24y_0^4 -6y_0^2 + 54y_0^4 -72y_0^6) \\
 & ~~~+ (-\tfrac{2}{3} + 14y_0^2 - 48y_0^4 + 40 y_0^6)
], \\
 = -T^2 &[ 
 ( -\tfrac{4}{3}        + 12y_0^2 - 36y_0^4 + 36 y_0^6)\\
 & ~~~+ (2              - 24y_0^2 + 78y_0^4 - 72 y_0^6) \\
 & ~~~+ (-\tfrac{2}{3}  + 14y_0^2 - 48y_0^4 + 40 y_0^6)
], \\
 = -T^2 &[ 2 y_0^2 - 6 y_0^4 + 4 y_0^6],\\
 = -T^2 &y_0^2(1-y_0^2)(1-2y_0^2)
\end{split}
\end{equation}
which matches the $T^2$ term of the true FTLE field \eqref{nonlinear_saddle_FTLE_approximation}.

\subsection{Details for the time-varying double-gyre example}\label{doublegyre_details}

The gradient tensor for the double-gyre velocity field \eqref{doublegyre} is,
\begin{equation}
\begin{split}
&\nabla\mathbf{v} =
\begin{bmatrix}
     \frac{\partial u}{\partial x} & \frac{\partial u}{\partial y} \\ 
     \frac{\partial v}{\partial x} & \frac{\partial v}{\partial y}  
\end{bmatrix}, \\
&=
\begin{bmatrix}
     -\pi^2 A\cos(\pi f)\cos(\pi y)\frac{\partial f}{\partial x} & 
      \pi^2 A\sin(\pi f)\sin(\pi y) \\ 
     -\pi^2 A\sin(\pi f)\sin(\pi y) \frac{\partial f}{\partial x} + \pi A \cos(\pi f)\sin(\pi y)\frac{\partial^2 f}{\partial x^2} & 
      \pi^2 A\cos(\pi f)\cos(\pi y)\frac{\partial f}{\partial x}  
\end{bmatrix}.
     \label{gradient_dg}
\end{split}
\end{equation}

The acceleration field, $\mathbf{a}=\tfrac{d}{dt}\mathbf{v}=(a_x,a_y)$, for the double-gyre, 
\eqref{doublegyre}, is given by,
\begin{equation}
\begin{split}
a_x =& -\pi^2 A  \cos(\pi f)\cos(\pi y)\tfrac{\partial f}{\partial t}
     +\tfrac{1}{2}\pi^3 A^2 \sin(2\pi f)\tfrac{\partial f}{\partial x}, \\ 
a_y =& ~~\pi^2 A \Big[ -  \sin(\pi f)\sin(\pi y)\tfrac{\partial f}{\partial x}\tfrac{\partial f}{\partial t}
     +\tfrac{1}{\pi}     \cos(\pi f)\sin(\pi y)\tfrac{\partial^2 f}{\partial x \partial t} \Big]
     \\
 +\tfrac{1}{2}&\pi^3 A^2  \sin(2\pi y) \Big[ \sin^2(\pi f) \tfrac{\partial f}{\partial x} + \cos^2(\pi f)( \tfrac{\partial f}{\partial x} )^2 -\tfrac{1}{2\pi}\sin(2\pi f)\tfrac{\partial^2 f}{\partial x^2}
      \Big],
\end{split}
\label{doublegyre_acc}
\end{equation}
where the dependence of the function $f$, from \eqref{f_function}, is understood.  


The components of the symmetric $\mathbf{B}$ matrix are,
\begin{equation}
\begin{split}
B_{xx} &= 
-A\pi^2 \cos(\pi f) \cos(\pi y) \tfrac{\partial^2 f}{\partial x \partial t}
+\tfrac{1}{2}A\pi^3 \sin(2 \pi f) \tfrac{\partial f}{\partial x}\tfrac{\partial f}{\partial t}\\
&
+A^2\pi^3 \sin(2 \pi f)\tfrac{\partial^2 f}{\partial x^2}
\Big( \tfrac{1}{2} - \sin^2(\pi y)(\tfrac{\partial f}{\partial x} )^2 \Big)
+
A^2\pi^4 \cos(2 \pi f)(\tfrac{\partial f}{\partial x} )^2 \\
&
+
A^2\pi^4 \sin^2(\pi f)\sin^2(\pi y)(\tfrac{\partial f}{\partial x} )^4
+
A^2\pi^2\cos^2(\pi f)\sin^2(\pi y) 
\tfrac{\partial^2 f}{\partial x^2}
 \\
&
+
A^2\pi^4\cos^2(\pi f)  \cos^2(\pi y) (\tfrac{\partial f}{\partial x} )^2,
\end{split}
\label{doublegyre_Bxx}
\end{equation}
\begin{equation}
\begin{split}
B_{xy} &=  
\tfrac{1}{2} A \pi \cos(\pi f) \sin(\pi y)
\Big[ \tfrac{\partial^3 f}{\partial x^2 \partial t}
+ \pi^2 \Big( 1 - (\tfrac{\partial f}{\partial x} )^2 \Big)
\Big]
\\
&
-A \pi^2 \sin(\pi f) \sin(\pi y)
\Big( \tfrac{\partial f}{\partial x}\tfrac{\partial^2 f}{\partial x \partial t} - \tfrac{1}{2}
\tfrac{\partial^2 f}{\partial x ^2}
\tfrac{\partial f}{\partial t}
\Big)
\\
&
-
\tfrac{1}{4}A^2 \pi^4
\sin(2 \pi f) \sin(2 \pi y)
\Big[
\tfrac{\partial f}{\partial x}
\Big( 1 + (\tfrac{\partial f}{\partial x} )^2 \Big)
+\tfrac{1}{2}\tfrac{\partial^3 f}{\partial x ^3}
\Big],
\end{split}
\label{doublegyre_Bxy}
\end{equation}
\begin{equation}
\begin{split}
B_{yy} &= 
A \pi^2 \cos(\pi f)  \cos(\pi y) 
\tfrac{\partial^2 f}{\partial x \partial t}
-
A \pi^3 \sin(\pi f) \cos(\pi y) \tfrac{\partial f}{\partial x} \tfrac{\partial f}{\partial t}
\\
&
-
\tfrac{1}{2}A^2 \pi^3 \sin(2 \pi f)\sin(2 \pi y)
\tfrac{\partial^2 f}{\partial x^2}
\\
&
+
A^2 \pi^4
\cos^2(\pi f) 
\cos^2(\pi y) (\tfrac{\partial f}{\partial x} )^2
+
A^2 \pi^4
\sin^2(\pi f) 
\sin^2(\pi y)
\\
&
+
A^2 \pi^4
(\tfrac{\partial f}{\partial x} )^2
\Big(
\cos^2(\pi f) 
- \sin^2(\pi y) 
 \Big).
\end{split}
\label{doublegyre_Byy}
\end{equation}
The eigenvalue $s_-(\mathbf{x}_0,t_0)$ of $\mathbf{S}(\mathbf{x}_0,t_0)$ is,
\begin{equation}
\begin{split}
s_- 
=
-\frac{1}{2}\pi^2 A \Bigg[  &
\Big(  \sin(\pi f) \sin(\pi y_0)\Big(1-\frac{\partial f}{\partial x} \Big) + \tfrac{1}{\pi}\cos(\pi f) \sin(\pi y_0)\frac{\partial^2 f}{\partial x^2}
\Big)^2 
\\
& ~~ +  4 \Big( \cos(\pi f) \cos(\pi y_0) \frac{\partial f}{\partial x} \Big)^2 
~\Bigg]^{1/2}.
\end{split}
\label{doublegyre_s1}
\end{equation}
The normalized eigenvector of $\mathbf{S}(\mathbf{x}_0,t_0)$ corresponding to the eigenvalue $s_-(\mathbf{x}_0,t_0)$ is given by,
\begin{equation}
  \bm{e}_{-}=
  \begin{bmatrix}
e_x \\
e_y
\end{bmatrix}
=
\frac{1}{N}  
\begin{bmatrix}
\bar s_- - \beta \\ 
\tfrac{1}{2}\alpha
\end{bmatrix}, \label{e1_doublegyre}
\end{equation}
where,
\begin{equation}
    \begin{split}
        \bar s_- &= \frac{s_-}{\pi^2 A} = -\tfrac{1}{2}
        \sqrt{\alpha^2 + 4\beta^2},\\
        N & = \sqrt{ \tfrac{1}{4}\alpha^2 + ( \bar s_1 - \beta)^2 }, \\
        \alpha & = \sin(\pi f) \sin(\pi y) \Big( 1 - \tfrac{\partial f}{\partial x} \Big) + \tfrac{1}{\pi} \cos(\pi f)\sin(\pi y)\tfrac{\partial^2 f}{\partial x^2}, \\
        \beta & = \cos(\pi f) \cos(\pi y)  \tfrac{\partial f}{\partial x}.
    \end{split}
\end{equation}
The coefficient of $T$ in the approximation of the backward-time FTLE for the double-gyre is thus given by
$s_-^2 - \tfrac{1}{2} \bm{e}_{-}^{\top} \mathbf{B} \bm{e}_{-}$  which can be expressed in terms of,
\begin{equation}
    a_-(\mathbf{x_0},t_0) = -s_-^2 +\tfrac{1}{2} (B_{xx} e_x^2 + 2 B_{xy} e_x e_y + B_{yy} e_y^2),
\end{equation}
using the above formulas. This
yields a backward-time FTLE approximation for small backward times $T<0$ of,
\begin{equation}
    \sigma_{t_{0}}^{t_0 + T}(\mathbf{x_0}) = s_-(\mathbf{x_0},t_0) - a_-(\mathbf{x_0},t_0)T + \mathcal{O}(T^2).
\end{equation}
Note that the first and second terms have explicit dependence on both initial position and initial time.

\subsection{Details for the ABC flow example}\label{ABCflow_example_details}
For the ABC velocity field \eqref{abcflow}, the characteristic polynomial for the rate-of-strain tensor $\mathbf{S}$ for this system is,
\begin{equation}
s^3 +a_1 s + a_0 = 0, \label{ABCflow_characteristic_polynomial}
\end{equation}
where 
\begin{equation}
\begin{split}
a_0 &= -\tfrac{1}{4}( B\cos(x)-C\sin(y))( C\cos(y)-A\sin(z))(-B\sin(x)+A\cos(z)),\\
a_1 &=-\tfrac{1}{4}\Big[ ( B\cos(x)-C\sin(y))^2 \\
&~~~~~~~~~~~~~ +( C\cos(y)-A\sin(z))^2 +(-B\sin(x)+A\cos(z))^2 \Big]
\end{split}
\end{equation}

The repulsion and attraction rate fields, $s_+$ and $s_-$, are given by
\begin{equation}
\begin{split}
s_+ &= 2 \rho^{1/3} \cos\big(\tfrac{\theta}{3}\big) >0, \\
s_- &= -\tfrac{1}{2}s_+ - \sqrt{3}\rho^{1/3}  \sin\big(\tfrac{\theta}{3}\big) <0,
\end{split}
\end{equation}
where 
the dependence on initial position $\mathbf{x}$ is understood and $\rho$ and $\theta$ are given by,
\begin{equation}
\begin{split}
\rho &= \sqrt{ q^2 + |p|}, \\
\theta &= \tan^{-1}\Big(\tfrac{{\rm Im}(\sqrt{p})}{q}\Big),
\end{split}
\end{equation}
where,
\begin{equation}
\begin{split}
q &=-\tfrac{1}{2}a_0,\\
p &=\frac{1}{27}a_1^3 + \frac{1}{4}a_0^2.
\end{split}
\end{equation}

\end{appendices}

\bibliographystyle{shane-unsrt} 
\bibliography{ref}

\end{document}